\documentclass[structabstract,natbib]{aa}  

\newcommand{\sse}{{\tt sourceExtractorSussextractor}}
\newcommand{\bnd}{{\tt sourceExtractorTimeline}}
\newcommand{\sex}{{\tt Sextractor}}
\newcommand{\hers}{{\it Herschel}}
\newcommand{\micron}{$\mu$m}
\usepackage{natbib}
\usepackage{graphicx}
\usepackage{txfonts}
\usepackage{array}
\usepackage{color}
\usepackage{rotating}
\bibpunct{(}{)}{;}{a}{}{,} % to follow A\&A style 

\begin{document}

\title{The {\it Herschel}\thanks{{\it Herschel} is an ESA space observatory with science instruments provided by a European-led principal investigator consortia and with an important participation from
NASA.} Virgo Cluster Survey}

\subtitle{XVII. SPIRE point-source catalogs and number counts}

\author{Ciro Pappalardo\inst{1,2}, George J. Bendo\inst{3}, Simone Bianchi\inst{4}, Leslie Hunt\inst{4}, Stefano Zibetti\inst{4}, Edvige Corbelli\inst{4}, Sperello di Serego Alighieri\inst{4}, Marco Grossi \inst{1,2},
Jonathan Davies\inst{5}, Maarten Baes\inst{6}, Ilse De Looze\inst{6}, Jacopo Fritz\inst{6}, Michael Pohlen\inst{5}, Matthew W. L. Smith\inst{5}, Joris Verstappen\inst{6},
M\'ed\'eric Boquien\inst{7}, Alessandro Boselli\inst{7}, Luca Cortese\inst{8,9}, Thomas Hughes\inst{6}, Sebastien Viaene\inst{6}, Luca Bizzocchi\inst{1,2}, Marcel Clemens\inst{10}}

\institute{
  Centro de Astronomia e Astrof\'{\i}sica da Universidade de Lisboa,  Observat\'{o}rio Astron\'{o}mico de Lisboa, Tapada da Ajuda, 1349-018 Lisbon, Portugal
  \and
  Department of Physics, Faculty of Sciences, University of Lisbon, Campo Grande, 1749-016 Lisbon, Portugal
    \and
  Jodrell Bank Centre for Astrophysics, Alan Turing Building, School of Physics and Astronomy, University of Manchester, Manchester M13 9PL
  \and
  Osservatorio Astrofisico di Arcetri - INAF, Largo E. Fermi 5, 50125 Firenze, Italy e-mail: {\tt cpappala@arcetri.astro.it}
  \and
  Department of Physics and Astronomy, Cardiff University, The Parade, Cardiff, CF24 3AA, UK
     \and
   Sterrenkundig Observatorium, Universiteit Gent, Krijgslaan 281 S9, B-9000 Gent, Belgium
  \and
  Laboratoire dÕAstrophysique de Marseille - LAM, Universit\'e d'Aix-Marseille \& CNRS, UMR7326, 38 rue F. Joliot-Curie, 13388 Marseille Cedex 13, France
   \and
   Centre for Astrophysics \& Supercomputing, Swinburne University of Technology, Mail H30 - PO Box 218, Hawthorn, VIC 3122, Australia
   \and
   European Southern Observatory, Karl-Schwarzschild Str. 2, 85748 Garching bei Muenchen, Germany
  \and
  Osservatorio Astronomico di Padova, Vicolo dell’Osservatorio 5, 35122 Padova, Italy
}

\date{}
\abstract
% context heading (optional)
    {} %leave it empty if necessary  
    % aims heading (mandatory)
    {We present three independent catalogs of point-sources extracted from SPIRE images at 250, 350, and 500\, \micron, acquired with the \hers\ Space Observatory as a part of the \hers\ Virgo Cluster Survey (HeViCS). The catalogs have been cross-correlated to consistently extract the photometry at SPIRE wavelengths for each object.}
    % methods heading (mandatory)
    {Sources have been detected using an iterative loop. The source positions are determined by estimating the likelihood to be a real source for each peak on the maps, according to the criterion defined in the \sse\ task. The flux densities are estimated using the \bnd, a timeline-based point source fitter that also determines the fitting procedure with the width of the Gaussian that best reproduces the source considered. Afterwards, each source is subtracted from the maps, removing a Gaussian function in every position with the full width half maximum equal to that estimated in \bnd. This procedure improves the robustness of our algorithm in terms of source identification. We calculate the completeness and the flux accuracy by injecting artificial sources in the timeline and estimate the reliability
      of the catalog using a permutation method.}
    % results heading (mandatory)
    {The HeViCS catalogs contain about 52000, 42200, and 18700 sources selected at 250, 350, and 500 \micron\ above 3$\sigma$ and are $\sim$75\%, 62\%, and 50\% complete at flux densities of 20\,mJy at 250, 350, 500, \micron, respectively.
       We then measured source number counts at 250, 350, and 500 $\mu$m and
      compare them with previous data and semi-analytical models. We also
      cross-correlated the catalogs with the Sloan Digital Sky Survey to investigate
      the redshift distribution of the nearby sources. From this cross-correlation,
      we select $\sim$2000 sources with reliable fluxes and a high signal-to-noise
      ratio, finding an average redshift $z$ $\sim$ 0.3 $\pm$ 0.22 and 0.25 (16-84 percentile)\thanks{The 250, 350, 500 \micron, and the total catalogs are available in electronic form at the CDS via anonymous ftp to cdsarc.u-strasbg.fr (130.79.128.5) or via http://cdsweb.u-strasbg.fr/cgi-bin/qcat?J/A+A/}.} 
    % conclusions heading (optional), leave it empty if necessary 
  {The number counts at 250, 350, and 500 $\mu$m show an increase in the slope below 200\,mJy, indicating a strong evolution in number of density for galaxies at these fluxes. In general, models tend to overpredict the counts at brighter flux densities, underlying the importance of studying the Rayleigh-Jeans part of the
      spectral energy distribution to refine the theoretical recipes of the models. Our iterative method for source identification allowed the detection of a family of 500 \micron\ sources that are not foreground objects belonging to Virgo and not found in other catalogs. 
      }
\titlerunning{Point Sources Catalogue}
\authorrunning{Pappalardo et al.}
\maketitle

%________________________________________________________________
\section{Introduction}

At least half of the energy emitted by the stars and active galactic nuclei (AGN) in a Hubble time has been absorbed by dust in the interstellar medium and re-emitted as IR
light \citep{pug,fix,lag2}.
The abundance and properties of dust are then crucial to determine the cosmic comoving star-formation rate density (SFRD), because the optical/ultraviolet emission originating from stars must be
corrected for IR extinction.
Moreover, if the bulk of far-IR (FIR) luminosity is due to the dust heated by massive stars in star-forming regions, the IR luminosity could be used as direct star formation tracer. However, at
$\lambda >$ 250 \micron, much of the dust emission in nearby galaxies may be dominated by dust heated by the total stellar population, mainly formed by evolved stars \citep[e.g., ][]{ben3,boq,ben2}.

Previous observations with IRAS \citep{sau,oli2}, ISO \citep{oli3}, and {\it Spitzer} \citep{shu,fra} have shown a strong evolution of galaxy luminosities in mid-IR and FIR wavelengths because of the decline
of the average SFRD with time. At redshift $z \sim$ 1, luminous IR galaxies contribute about 50\% to the SFRD \citep{murphy11}, while the SFRD increases at higher redshift, $z \sim 2$ because we have
an important contribution from more active star-forming galaxies, or ultraluminous IR galaxies \citep[$L_{IR} \ge 10^{12}$ $L_\odot$, ][]{cha,dad,pap,mag,mag2}, which are considered to be the ancestors of massive quiescent galaxies in the present epoch \citep[e.g.][]{lef}.

To determine the contribution of IR light from less luminous galaxies to the SFRD, deep IR observations are needed. Previous works have shown that confusion is the most important component of noise at
FIR wavelengths \citep{bla}. Faint unresolved sources are collected inside the observing beam, and the sky background in the
resulting map is produced mostly by these objects. The bulk of this unresolved emission consists of dusty star-forming galaxies at a redshift $z \ge$ 1.2 \citep{dev} and forms the FIR background
\citep[FIRB, ][]{pug,fix,lag2}.

The \hers\ Space Observatory \citep{pil} with its two spectro-photometers, SPIRE \citep{gri} and PACS \citep{pog}, has the necessary sensitivity to investigate the FIRB in greater detail, directly measuring
the IR luminosity function at higher redshift. The starting point for such an investigation is the construction of a list of unresolved sources down to the confusion noise limit.
These data can be used to infer statistical properties of galaxy populations to determine their evolution. A tool widely used to investigate galaxy evolution over Hubble time is the number density of
galaxies as a function of flux density, or the number counts. They are a direct tracer of the past galaxy evolution that can be compared to theoretical models.

\begin{figure} 
\begin{center} 
\includegraphics[clip=,width=.49\textwidth]{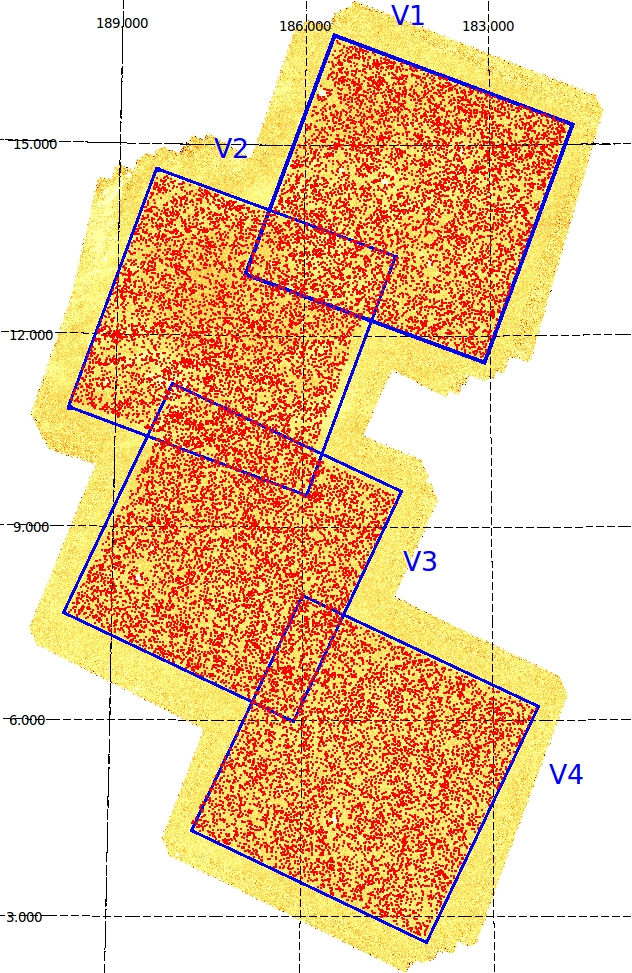} 
\end{center} 
\caption{HeViCS image at 250 $\mu$m with detected sources (red points). The blue squares indicate regions with the full coverage.}
\label{covfig} 
\end{figure}

In this paper, we present three catalogs of point-sources extracted from the \hers\
Virgo Cluster Survey \citep[HeViCS,][]{dav}, a survey that observed a region that is about 84 square degrees centered in the Virgo cluster, of which 55 square degrees 
are observed at uniform depth. In this survey, \hers\
observed with SPIRE and PACS four overlapping fields in fast parallel-mode, labeled as V1, V2, V3, and V4 in Fig. \ref{covfig}, which are 16 square degrees each \citep[see][for details about the survey and
the data reduction]{aul}. In this paper we considered only SPIRE data in the analysis, leaving the PACS catalog and data analysis to a future companion paper (Fritz et al, in prep.).

Although the primary goal of HeViCS was to study the galaxies within the Virgo Cluster itself, the survey serendipitously detected a substantial number of background galaxies, making it an excellent
survey for studying the FIRB. 

The paper is organized as follows: in Sec.
\ref{method}, we describe the method for the determination of the source flux and position and show how we estimate the background. 
We also investigate the goodness of our catalog through the
reliability, the completeness, and the accuracy in flux estimation. 
In Sec. \ref{analisi}, we discuss the number counts of the HeViCS catalogs with an analysis of the redshift distribution of the sources, which are obtained by
cross-correlating the catalog with Sloan Digital Sky Survey catalog \citep[SDSS - DR10,][]{ahn}. Our conclusions are given in Sec. \ref{conc}.

%----------------------------------------------------------------------------------------------------------
%----------------------------------------------------------------------------------------------------------

%-----------------------------------------------------------------------------------------------------------
\begin{table}\caption[]{Regions fully sampled, which are selected in each HeViCS field for the source extraction with flux densities $>$ 20\,mJy.  
Column 1: field; column  2, 3: center coordinates; column 4: position angle (counterclockwise); column 5: area
covered; column 6: number of sources
detected.}\label{covtable}
\begin{center}
\begin{tabular}[0.8\textwidth]{cccc >{\centering\arraybackslash}p{1.2cm} c}
\hline \hline
Field  & RA & DEC & Pos. Angle & Area & 250 $\mu$m \\ 
& deg & deg & deg & deg$^2$ & sources\\ 
(1) & (2) & (3) & (4) & (5) & (6)\\ 
\hline\\ 
V1 & 184.3 & 14.15 & 20 & 4 $\times$ 4 & 14061\\ 
V2 & 187.2 & 12.07 & 20 &4 $\times$ 4 & 12956\\ 
V3 & 187.1 & 8.6 & 25 & 4 $\times$ 4 & 13828\\ 
V4 & 185.0 & 5.25 & 20 & 4 $\times$ 4 & 14494\\ 
Combined & $-$ & $-$ & $-$ & 55 & 52020\\ 
\hline \hline
\end{tabular}
\end{center}
\end{table}
%-----------------------------------------------------------------------------------------------------------

%%%%%%%%%%%%%%%%%%%%%%%%%%%%%%%%%%%%%%%%%%%%
\section{Methodology} \label{method}

The determination of the position and the flux density of the sources have been performed in different steps. From the starting map, presented in Sec. \ref{data}, an estimate of the background has been subtracted (Sec. \ref{back}). The source positions have been determined from this background-subtracted map using the \sse\ software. Finally, the flux density in each position was estimated using the \bnd\ software, and each source has been subtracted from the timeline data once determined its flux density (Sec. \ref{identify}).

\subsection{Data}
\label{data}

The four HeViCS fields have been observed in two orthogonal scans at a sampling rate of 60$''s^{-1}$ with four repetitions for a total of eight scans. The final map is obtained by averaging the
bolometer timeline measurements that sampled the region in each map pixel \citep[{\tt naiveScanmapper} task in HIPE 9.0.0,][]{ott}, as described in detail in \cite{aul}. The main difference between our pipeline and the standard one is that we did not run the default {\tt temperatureDrift-Correction} task and the residual, median baseline subtraction. Instead we use the {\it BRIght Galaxy ADaptive Element} (BriGAdE, Smith et al., in prep.) which uses an alternative technique for correcting temperature drifts. The SPIRE instrument
\citep{gri}{\footnote{\it{http://herschel.esac.esa.int/Docs/SPIRE/pdf/spire\_om.pdf}}} has three bolometer arrays that observed the sky at 250, 350, and 500 $\mu$m. The SPIRE calibration tree v8 was used and the full width at half-maximum
(FWHM) measurements of the beams in the final maps for point sources are 17 \farcs5, 23\farcs9 and 35\farcs1 at 250, 350, and 500\,\micron, respectively{\footnote{{\it http://herschel.esac.esa.int/twiki/pub/Public/SpireCalibrationWeb\\ /beam\_release\_note\_v1\-1.pdf} and {\it http://herschel.esac.esa.int/twiki/bin\\
/view/Public/SpirePhotometerBeamProfileAnalysis}}}.  The maps we used have pixel sizes of 6\arcsec, 8\arcsec, and 12\arcsec\ at 250, 350, and 500\,\micron, respectively. The best spatial resolution and the deepest data are at 250 $\mu$m, and for this reason, we chose the source positions at this wavelength as a reference for HeViCS catalogs.

\subsection{Background subtraction} \label{back}

\begin{figure}
\begin{center}
\includegraphics[clip=,width= .49\textwidth]{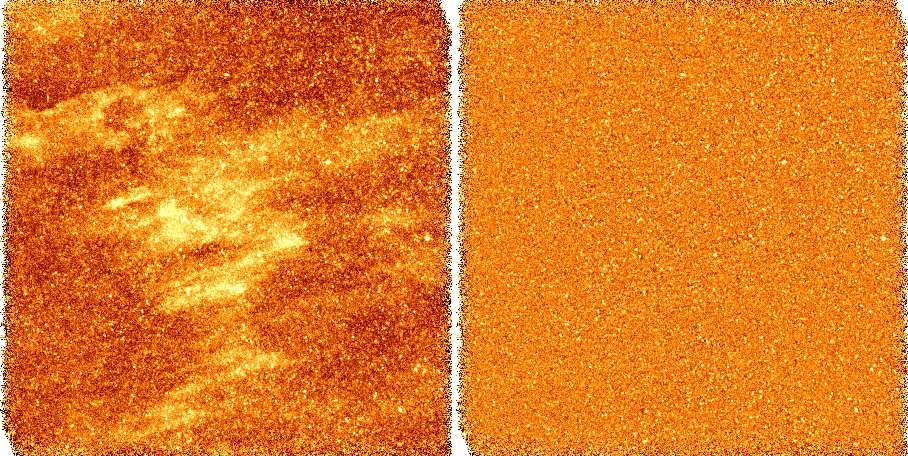}
\end{center}
\caption{V2 field region before (left panel) and after (right panel) background subtraction.}
\label{bckgsub}
\end{figure}

The SPIRE data are contaminated by Galactic cirrus emission, particularly in the V2 field. The cirrus emission does not significantly affect the flux densities measured using the timeline-based source
fitter, but it does cause \sse\ to identify more sources in regions with strong cirrus emission. This is because the diffuse component in the convolved map is considered by \sse\ as a sequence of
peaks with each one a high likelihood to be a source (see Sec. \ref{identify} for details). To mitigate this effect, we estimate the background using a method developed in \sex\ \citep{ber}. In this
method, the map is regridded into cells larger than the pixel size, and then we iteratively estimate the mean and the standard deviation of the distribution of pixel values in each cell. At each step, the
outlier pixels are removed until the variation of $\sigma$ in an iteration is small \citep[see][ for details]{ber}. The value of the background in each block is then estimated by using 2.5 $\times$
$M_{med} - 1.5 \times M_{avg}$, where $M_{med}$ and $M_{avg}$ are the median and the mean inside each block. Finally, the background map is obtained by linearly interpolating between the cells. The
size of the blocks fixes the size of the structure that we are able to remove. If the cell size is too large, it results in a map in which the diffuse emission is not well constrained, while a cell
size that is too small results in a map that reproduces the diffuse emission well but does not remove the extended emission due to the resolved extragalactic sources larger than the beam. After
performing a series of tests, we chose to fix the block size to 8 pixels, slightly smaller than the beam size (9 pixels). Figure \ref{bckgsub} shows a region of the V2 field before and after cirrus
subtraction.  This background is subtracted from the map data that is used as input by \sse\ in the source identification step described in Section~\ref{identify}.

%-------------------------------------------------------------------------------------------------
%-----------------------------------------------------------------------------------------------------------

%-----------------------------------------------------------------------------------------------------------
\begin{figure}\begin{center}
\includegraphics[clip=,width= .49\textwidth]{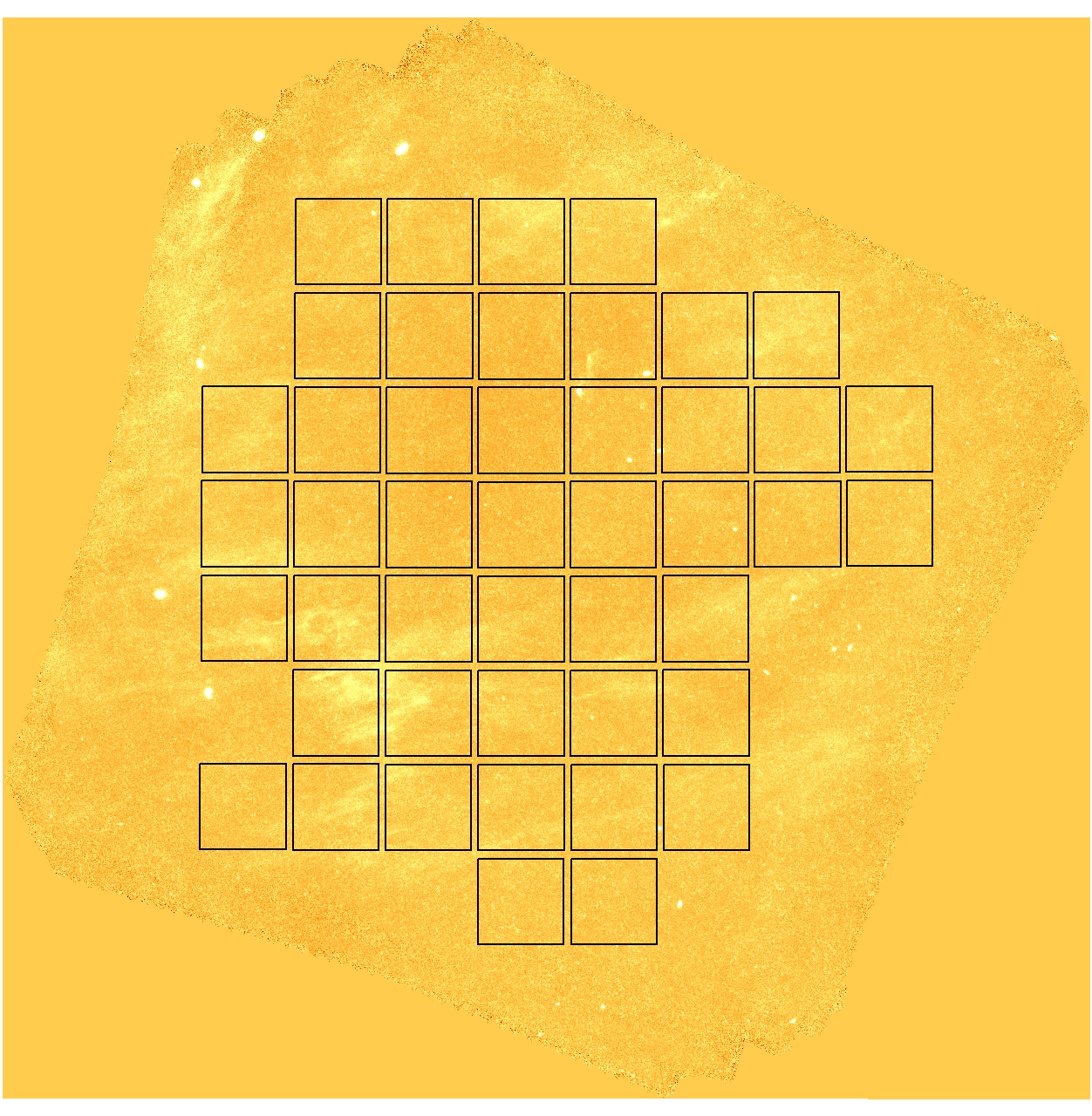}
\includegraphics[clip=,width= .49\textwidth]{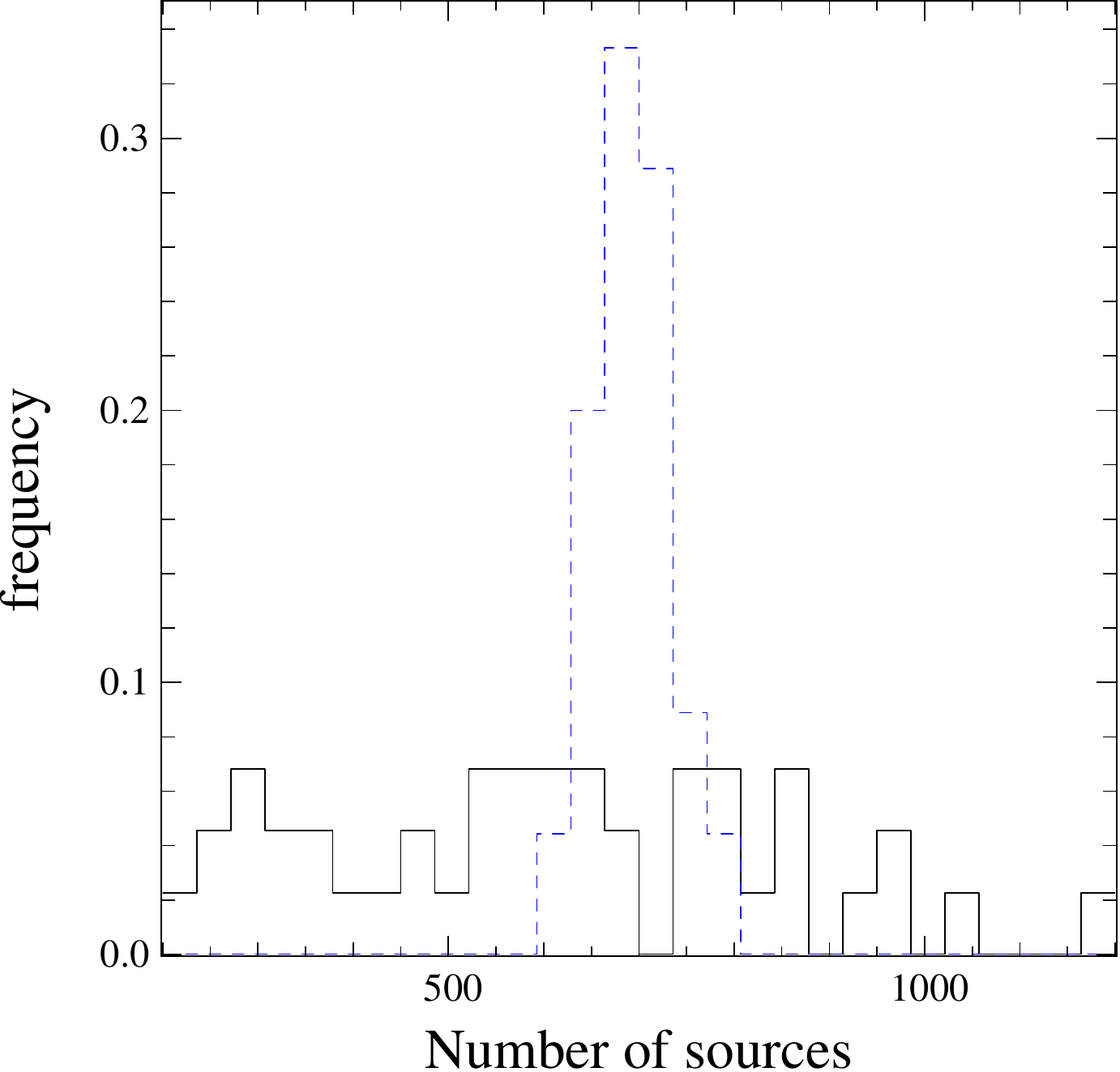}
\end{center}\caption{Top panel: Boxes used in V2 field to test cirrus subtraction. Bottom panel: Distribution of the number of 
sources detected in the V2 boxes shown in the top panel before (black solid line) and after (blue dashed line) background subtraction.}\label{55boxes}\end{figure}
%-----------------------------------------------------------------------------------------------------------

To test the reliability of this background estimation, we selected 45 boxes of 0.5$\times$0.5\,deg$^2$ that cover different regions in the V2 field with strong and weak cirrus emission (top panel of
Fig. \ref{55boxes}). In each box, we estimate the number of sources detected before and after the background subtraction. We fix a low threshold for the source-detection ($\sim$ 5 mJy) because we
want to also consider the sources with flux densities well below the limit chosen for the catalog for this test. Selecting a higher threshold would introduce a bias due to sources with flux densities around the
detection limit. If the background subtraction has worked properly, we should measure the same number of sources in each box with relatively small variations in the total number of sources detected
in each box. Results are shown in bottom panel of Fig \ref{55boxes}. Before the subtraction, the number of sources detected with \sse\ spans a high range of values, 188 $\le N_{bef} \le$ 1199 with an
average value of 600 $\pm$ 242.7 (bottom panel of Fig. \ref{55boxes}). After the background subtraction the number of sources is quite homogeneous for regions with either strong or
weak cirrus contamination with an average value of 697 sources and a significantly lower standard deviation of 37.9 (blue dashed line in bottom panel of Fig. \ref{55boxes}). The number of detections above 20 mJy are consistent with the results obtained with the previous test. They reproduce a similar trend but are rescaled, because of the smaller number of sources due to the higher threshold. The background-subtraction process reduces the standard deviation of the detections above 20 mJy from 50\% (before subtraction) to 10\% (after the subtraction). This reduction in the standard deviation and the homogeneity of the number of detections after the background subtraction confirm the effectiveness of our cirrus removal methods.

\subsection{Source identification and flux density measurement} \label{identify}

Source detection is performed in an iterative loop. A list of initial positions of point sources is produced using the implementation of \sse\ \citep{sav} in HIPE 9.0.0 \citep{ott} that is applied to a map where the foreground cirrus structure has been
subtracted (see Section~\ref{back}). The \sse\ software convolves the reduced map with a bidimensional Gaussian kernel having a FWHM of 17\farcs5, 23\farcs9, and 35\farcs1 at 250, 350, and 500 $\mu$m, respectively. The value of each pixel in the convolved map corresponds to the maximum likelihood estimate of the flux density that a source should have if centered on that pixel. The extractor routine finally searches for local maxima in pixel
regions with a size comparable to the beam, which are the neighboring eight pixels. For this analysis, we only used sources from locations for which we have a redundancy of $\geq$\,35 bolometer samplings per
pixel, or eight coverages (see Table \ref{covtable} and Fig. \ref{covfig}). Since the main source of noise in our maps is the confusion given by the FIRB \citep{pug},  we only use sources with
initial flux density measurements of $>$ 20\,mJy. This threshold corresponds approximately to 3$\sigma$ above the confusion noise \citep[estimated in $\sim$ 6-7\,mJy beam$^{-1}$ at 250, 350, and 500 \micron,][]{ngu}.

Each source is then fitted with a Gaussian function using \bnd, a timeline-based point source fitter \citep[see][]{ben} implemented in HIPE 9.0.0 \citep{ott}. The method fits timeline data from all bolometers within an individual array with a two-dimensional Gaussian function.  As an input, the fitter uses the source position given by \sse, an aperture containing the target data (using apertures of 22\arcsec, 30\arcsec, and 42\arcsec\ at 250, 350, and 500~$\mu$m, respectively), and an aperture identifying the background annulus (set to a radius of 700-720\arcsec ). The output consists of parameters describing the amplitude, position, the FWHM, and the parameters describing the background flux density. 

Two fits are performed to the timeline data.  We first fit a circular Gaussian function in which the FWHM is allowed to vary.   We then use the FWHM to determine whether the source is resolved or unresolved and reject sources that are either too narrow (which may be unremoved glitches) or too broad (which are probably extended sources). To test at which distance two sources are distinguishable, we injected in the timeline data point-like objects with the same flux density and an increasing distance from each other. We found that two sources are resolved only if they are at a distance above $\sim $22\arcsec, 30\arcsec and 46\arcsec at 250, 350, and 500 \micron, respectively. Fixing the distance between the sources to these values, we then injected in the timeline data point-like objects at different flux densities to investigate the FWHM recovered by \bnd. Based on these tests, we have found that artificial sources with FWHM corresponding to the telescope beam (17\farcs5, 23\farcs9, and 35\farcs1 at 250, 350, and 500 $\mu$m, respectively{\footnote{\it http://herschel.esac.esa.int/twiki/pub/Public/SpireCalibrationWeb\\ /beam\_release\_note\_v1\-1.pdf}}) added to timeline data may have FWHM between 10\arcsec\ and 30\arcsec\ at 250~$\mu$m, 13\farcs3 and 40\arcsec\ at 350~$\mu$m, and 20\arcsec\ and 60\arcsec\ at 500~$\mu$m. Sources, whose FWHM are determined via \bnd, and were outside these values have been excluded from the analysis.

The flux densities of resolved sources in this range of FWHM are calculated fitting a circular Gaussian function with the FWHM set to the measured SPIRE beam; we explain how this produces the most reliable flux densities for our targets in Section~\ref{sec:completeness}.

Although \sse\ produces flux density measurements for the sources it detects and although other software packages have been developed for source extraction within confused extragalactic fields
(including software developed specifically for {\it Herschel}), we prefer to use the timeline-based source extraction for several reasons. First of all, the SPIRE data are flux calibrated at the timeline level using timeline-based PSF-fitting techniques \citep{ben}, so the methods that we have used are consistent with the flux calibration measurements themselves and are therefore expected to be more accurate.  Second, when the maps are produced from the timeline data, multiple measurements from the timelines are averaged together to produce the surface brightness measured in each pixel. The mapmaking process effectively leads to the loss of information about the emission from the source.  Because of this, it is preferable to use the timeline data for performing photometry
measurements through PSF fitting down to a level of 30 mJy, as the timeline data preserves the surface brightness and positional information of each individual measurement \citep[see also,][ and their tests with different methods]{pea}.  We also note that \sse\ itself is subject to some systematic bias issues related to how it uses the error map for weighting \citep{smi}.

Once we calculated the flux densities of each resolved source, we removed them from the timeline data using the flux density and the width of the Gaussian obtained in the fit.  This subtraction step removes bright sources from the data that may either cause problems when fitting fainter sources or possibly hide other objects. We repeat the source identification with \sse, the source fitting with \bnd, and the source subtraction for a total of three times on the data.  In the second and third iterations, we work with the images where we had subtracted all the sources identified in previous iterations.  This procedure increases the number of sources detected by $\approx$ 10\%, as it allows us to detect sources that may be missed in just one iteration.

\subsection{Extended sources}

Another problem with our method is that \sse\ tends to break up extended objects (generally Virgo cluster galaxies) into ``families'' of point sources, which could potentially contaminate our
catalog. To address this problem, we performed a series of tests to determine how to remove bright extended sources a priori.  The method we developed defines a mask using the approach of \sex\
\citep{ber}, which detects a source when a fixed number of contiguous pixels is above a $\sigma$-threshold estimated from the background map. We estimate that 70 contiguous pixels above 1.2$\sigma$ is
a reasonable threshold to remove most of extended sources. We therefore reject sources with a total of 250\,\micron\ flux density higher than $\sqrt{70} \cdot 1.2 \approx 10\sigma$ $\approx$
60\,mJy and larger than 0.7\,arcmin$^2$; we reject all sources detected by \sse\ that conform to these criteria. In this sense, these maps are not really masks in the usual sense of the words but a way to define regions where the sources are rejected.

%--------
\subsection{Quality assessment} \label{quality}

In this section, we determine the reliability and the completeness of the catalogs with different tests. Then, we also verify the flux accuracy in
the \bnd\ fitter, and the precision in the determination of the source position
in \sse.

\subsubsection{Reliability}

The reliability of a catalog can be quantified by the number of false
detections found at a given flux density. When the dominant sources of noise are
the background fluctuations, a false detection is simply a source that has a
nominal signal-to-noise ratio below a fixed threshold. However, when the noise is dominated by the confusion, or the number of 
faint unresolved sources inside
a beam is high, such a criterion is no longer valid. In a crowded field,
the estimated flux density is uncertain, since each detection can be
contaminated by neighboring sources.

To quantify the number of false detections due to the instrumental noise, we use a permutation method similar to that used by \cite{smi}. We build two maps using the two halves of the eight
HeViCS-scans and calculate two difference maps:

\begin{equation}
M_{diff} = \pm \frac{M_1 - M_2}{2},
\label{rel}
\end{equation}   

\noindent
where $M_1$ and $M_2$ are the maps built with the first and the last four
scans. This procedure removes the confusion noise and considers only the uncertainties
due to the instrument. The subtraction is done in two different permutations
because a cosmic ray can give a positive or negative spike depending on the map
considered. 
We search for sources using \sse, which found 13, 21, 3, and 8 sources in V1, V2, V3, and V4, respectively. However the position of the sources are confined
to the external regions of the map in all cases, where the number of bolometer samplings is
low. In the catalog, we considered only regions of the field with a sampling
redundancy $\geq$\,35, therefore ignoring all the sources on the borders. This
implies that we can reasonably assume that there are no false detections due to glitches or different noise properties in different scans in the regions considered for the source extraction.  

\subsubsection{Completeness and flux accuracy}
\label{sec:completeness}

To determine the completeness and the flux accuracy of the catalog, we artificially added a Gaussian source at a random position to the timeline data with a FWHM equal to the beam for the appropriate photometric filter.  
In each HeViCS field, we define a box of 0.5$\times$0.5\,deg$^2$
and add a source with flux densities of 5, 10, 20,
30, 40, 50, 60, 70, 80, 90, and 100\,mJy. For each flux value, the test is
repeated 500 times.

We then extract a list of point sources in the selected box using \sse\
and search for the nearest source at a distance equal to at most half 
the beam. We avoid a larger distances because the fields are crowded, and
we can easily find an unrelated source close to the input coordinate. 

%-----------------------------------------------------------------------------------------------------------
\begin{table}\caption[]{Test of completeness and flux accuracy at 250 $\mu$m. Column 1: field; column 2: input flux in mJy; column 3, and 4: percentage of sources detected and not detected; column 5:
percentage of statistical outliers; column 6, 7, and 8: average, median, and standard deviation of the flux densities recovered, respectively, excluding the statistical outliers.}\label{V1comp}
\begin{center}
\begin{tabular}{ccccc|ccc}
\hline
 Field& F$_{inp}$& N$_{det}$& N$_{ndet}$& N$_{out}$ & $F_{avg}$ & $F_{med}$ & $F_{err}$ \\
  & mJy & \% & \% & \% & mJy & mJy & mJy \\
   (1) &  (2) & (3) & (4) & (5) & (6) & (7) & (8)\\ 
 &&&& \\ 
  V1  &&&&\\
 &      100 & 99.0 & 1.0 & 0.8 & 100 &  99 &   6 \\ 
 &       90 & 99.6 & 0.4 & 2.2 &  90 &  89 &   5 \\ 
 &       80 & 99.2 & 0.8 & 2.2 &  80 &  79 &   5 \\ 
 &       70 & 99.4 & 0.6 & 2.2 &  70 &  69 &   5 \\ 
 &       60 & 98.8 & 1.2 & 0.6 &  61 &  59 &   7 \\ 
 &       50 & 97.8 & 2.2 & 2.2 &  50 &  50 &   5 \\ 
 &       40 & 94.8 & 5.2 & 2.4 &  40 &  40 &   6 \\ 
 &       30 & 87.2 & 12.8 & 2.2 &  30 &  29 &   5 \\ 
 &       20 & 75.0 & 25.0 & 1.6 &  21 &  20 &   5 \\ 
 &       10 & 32.0 & 68.0 & 1.0 &  17 &  16 &   8 \\ 
 &        5 & 13.2 & 86.8 & 0.2 &  23 &  16 &  19 \\ 
 V2  &&&&\\
 &      100 & 100.0 & 0.0 & 1.6 & 100 & 100 &   6 \\ 
 &       90 & 100.0 & 0.0 & 1.8 &  90 &  90 &   5 \\ 
 &       80 & 99.4 & 0.6 & 1.0 &  81 &  80 &   6 \\ 
 &       70 & 99.8 & 0.2 & 1.4 &  70 &  69 &   6 \\ 
 &       60 & 98.8 & 1.2 & 2.2 &  60 &  60 &   6 \\ 
 &       50 & 97.2 & 2.8 & 1.8 &  50 &  49 &   5 \\ 
 &       40 & 95.2 & 4.8 & 1.6 &  41 &  40 &   6 \\ 
 &       30 & 91.8 & 8.2 & 1.0 &  31 &  30 &   6 \\ 
 &       20 & 75.8 & 24.2 & 1.6 &  21 &  21 &   5 \\ 
 &       10 & 40.4 & 59.6 & 1.4 &  15 &  14 &   6 \\ 
 &        5 & 16.0 & 84.0 & 0.2 &  19 &  14 &  16 \\ 
 V3  &&&&\\
 &      100 & 100.0 & 0.0 & 1.2 & 100 &  99 &   5 \\ 
 &       90 & 99.8 & 0.2 & 2.0 &  90 &  90 &   6 \\ 
 &       80 & 99.8 & 0.2 & 2.6 &  80 &  79 &   6 \\ 
 &       70 & 100.0 & 0.0 & 1.4 &  70 &  70 &   6 \\ 
 &       60 & 98.4 & 1.6 & 2.0 &  60 &  59 &   6 \\ 
 &       50 & 98.0 & 2.0 & 2.4 &  50 &  49 &   5 \\ 
 &       40 & 95.2 & 4.8 & 1.6 &  40 &  39 &   6 \\ 
 &       30 & 84.2 & 15.8 & 1.4 &  31 &  30 &   6 \\ 
 &       20 & 68.8 & 31.2 & 1.6 &  22 &  21 &   6 \\ 
 &       10 & 35.6 & 64.4 & 0.6 &  19 &  16 &  11 \\ 
 &        5 & 13.4 & 86.6 & 0.2 &  23 &  16 &  28 \\ 
 V4  &&&&\\
 &      100 & 100.0 & 0.0 & 2.0 & 100 &  99 &   6 \\ 
 &       90 & 100.0 & 0.0 & 2.2 &  89 &  89 &   5 \\ 
 &       80 & 99.4 & 0.6 & 1.8 &  80 &  80 &   5 \\ 
 &       70 & 99.8 & 0.2 & 1.2 &  70 &  69 &   5 \\ 
 &       60 & 99.6 & 0.4 & 2.0 &  60 &  59 &   6 \\ 
 &       50 & 98.8 & 1.2 & 1.8 &  50 &  49 &   6 \\ 
 &       40 & 96.4 & 3.6 & 1.4 &  40 &  39 &   5 \\ 
 &       30 & 90.8 & 9.2 & 1.6 &  30 &  30 &   5 \\ 
 &       20 & 77.4 & 22.6 & 1.0 &  22 &  21 &   6 \\ 
 &       10 & 37.8 & 62.2 & 0.2 &  19 &  16 &  14 \\ 
 &        5 & 17.6 & 82.4 & 1.2 &  19 &  15 &  14 \\ 
\hline
\end{tabular}
\end{center}
\end{table}
%-----------------------------------------------------------------------------------------------------------

\begin{figure*}\begin{center}
\includegraphics[clip=,width= .33\textwidth]{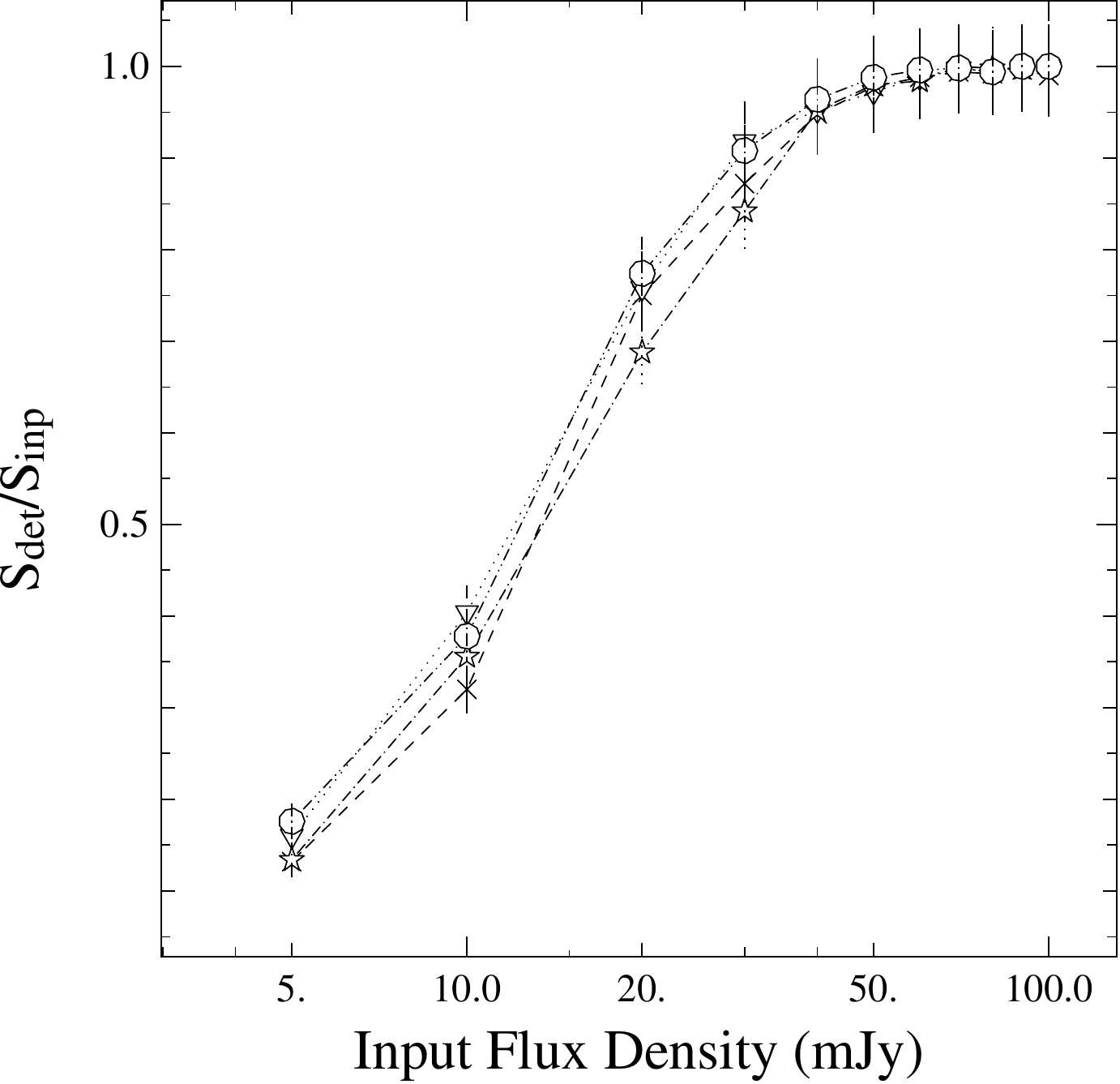}
\includegraphics[clip=,width= .33\textwidth]{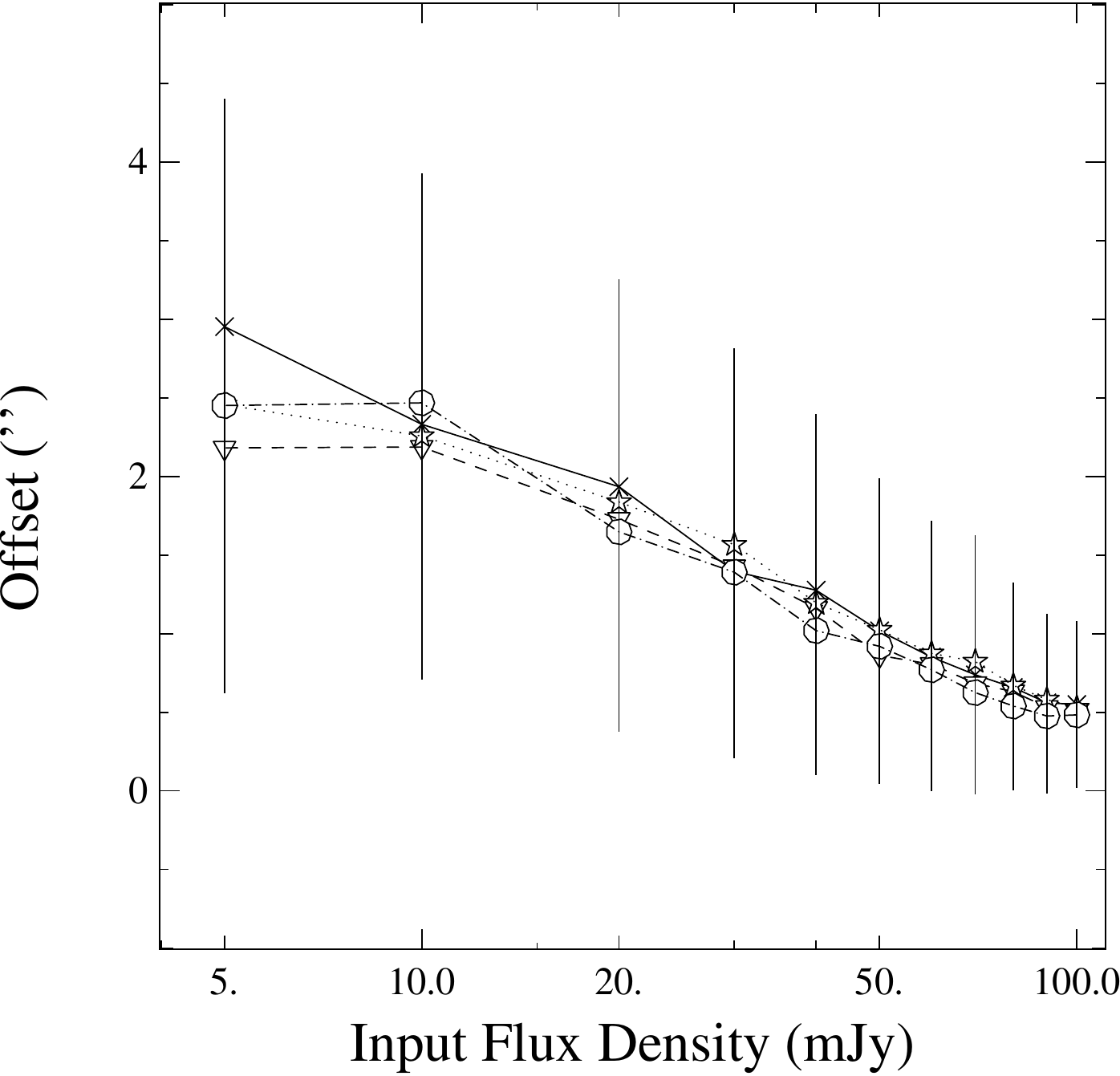}
\includegraphics[clip=,width= .33\textwidth]{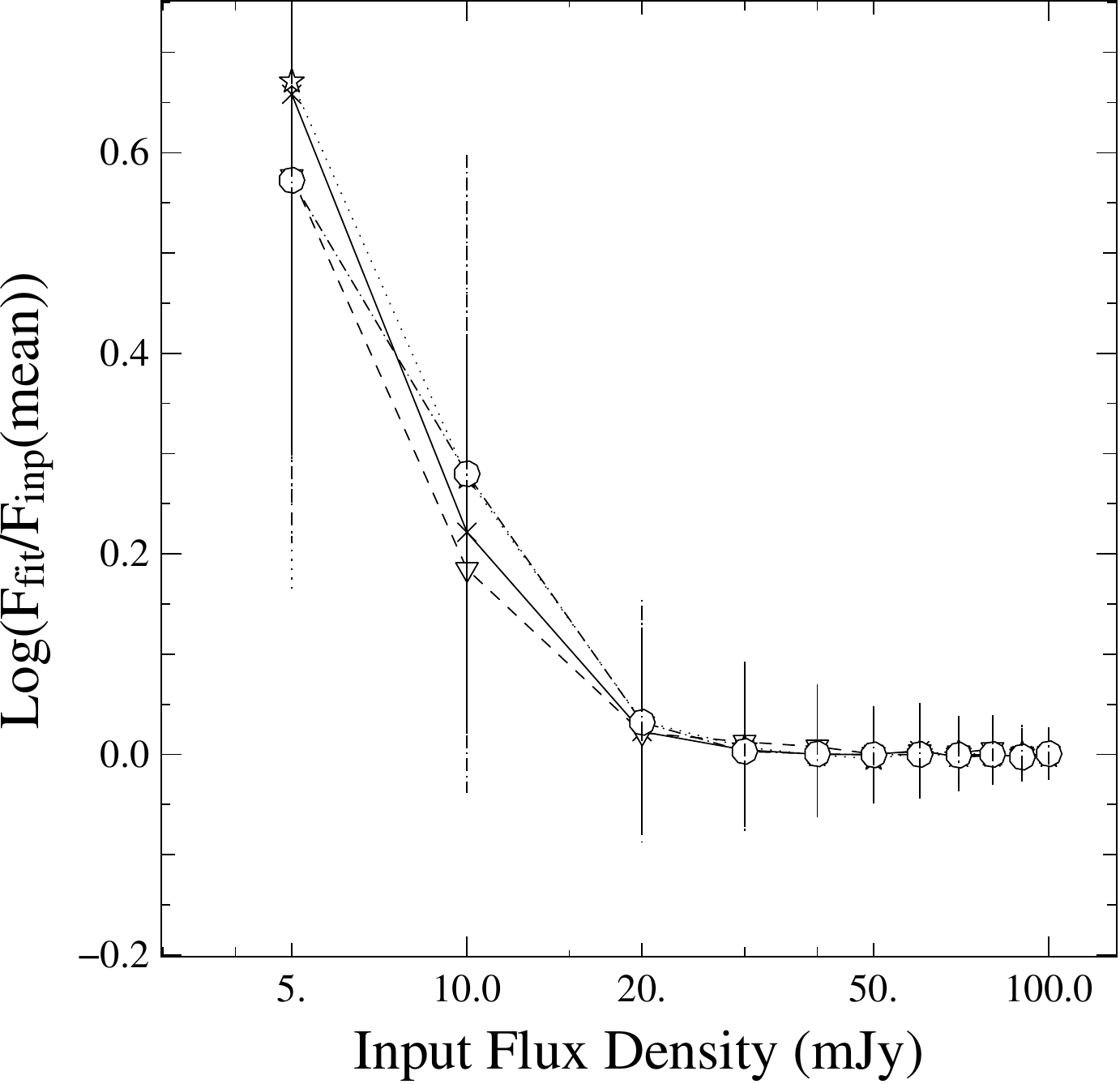}
\end{center}
\caption{Completeness and flux-position accuracy in the catalog at 250 $\mu$m. In each panel, the field V1, V2, V3, and V4 are shown as solid, dashed, dotted, and dash-dotted line. Left panel: Ratio
between the numbers of objects detected and the number of sources injected. Middle panel: Radial offset between the sources injected and the source recovered as a function of the
input flux density. Right panel: Logarithm of the ratio between the average recovered flux density and the input one, as a function of the injected fluxes. 
}
\label{randomsource}
\end{figure*}

As a preliminary test, we verified that the completeness and flux accuracies are consistent all over for the four HeViCS fields. Table \ref{V1comp} and Fig. \ref{randomsource} show the results of this test at 250 $\mu$m: columns 1 and 2 indicate the field selected and the flux
density of the source added in the timeline; columns 3 and 4 give the
percentage of sources detected and not detected; and column 5 reports the statistical outliers, defined as the sources with flux densities that differ by more than 3$\sigma$ from the measured average. In our case, this value ranges from 0.2\% to 2.2\%, indicating that the flux density calculated with the \bnd\ is robust when the source is detected.

The fraction of high-flux sources detected at 250 \micron\ has more than 95\% of the sources that are down to 40\,mJy, independent of the chosen fields (left panel of Fig. \ref{randomsource}). Below this value, the completeness
level decreases rapidly to $\sim$ 88\% at 30\,mJy and to $\sim$ 74\% at 20\,mJy, as expected by considering the confusion noise level of $\sim$ 20\,mJy at 3$\sigma$ found by
\cite{ngu}. The flux estimation of \bnd\ is quite stable until 20\,mJy, despite the rapid decrease of the completeness to 74\%, where the recovered flux is
on average 21$\pm$5\,mJy, and agrees with the input value (right panels of Fig.
\ref{randomsource}). However, below this threshold, the fluxes of sources
between 5 and 10\,mJy are systematically overestimated because the effect called ``flux boosting'' becomes relevant at these values.  
This last result also confirms that we find only a ``sea'' of faint sources, which constitutes the unresolved far-IR background, below a flux density of $\sim$ 20\,mJy. 

%----------------------------------------------------------------------------------------------------------
%----------------------------------------------------------------------------------------------------------
%----------------------------------------------------------------------------------------------------------
\begin{figure*}\begin{center}
\includegraphics[clip=,width= .33\textwidth]{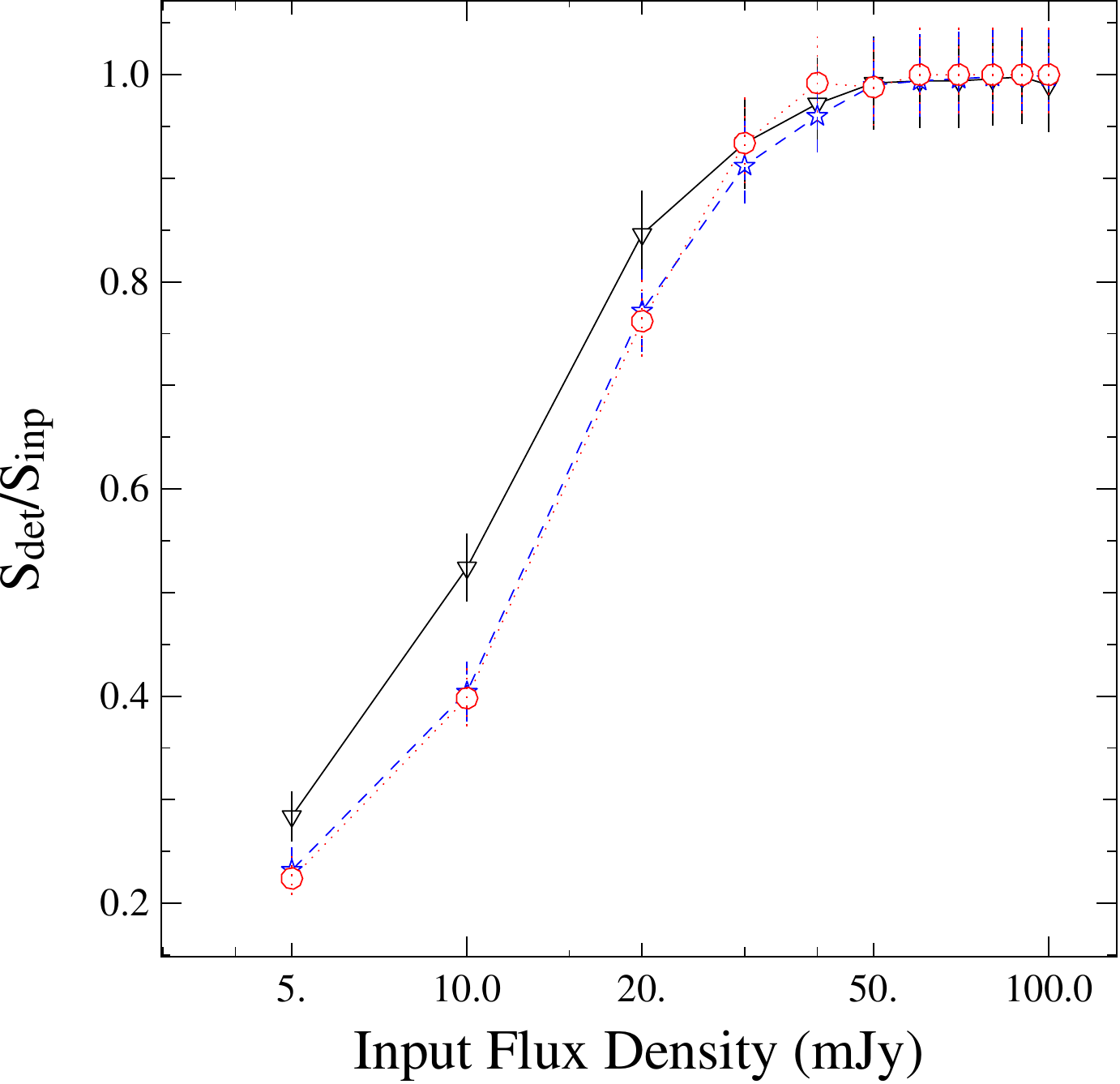}
\includegraphics[clip=,width= .33\textwidth]{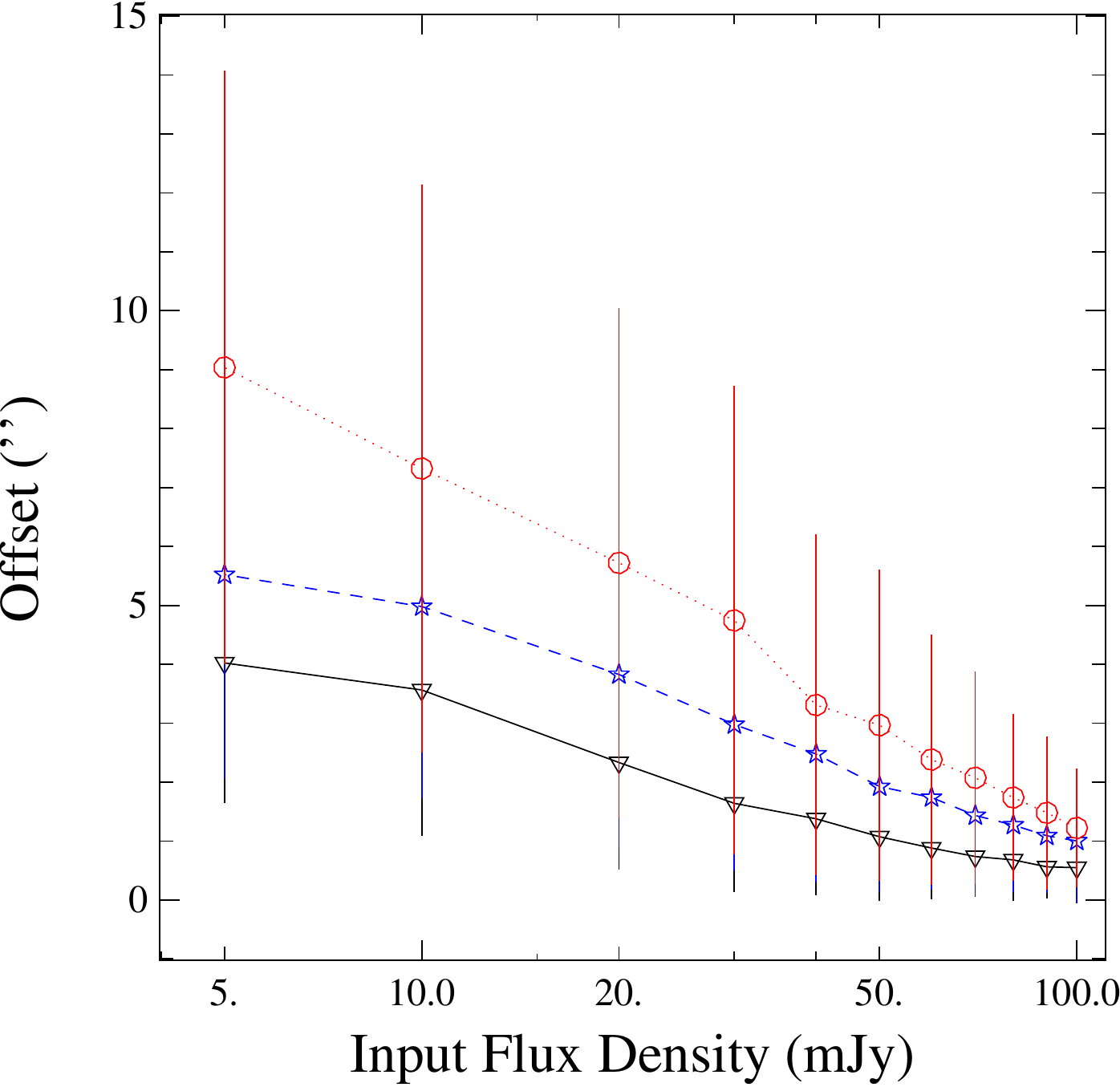}
\includegraphics[clip=,width= .33\textwidth]{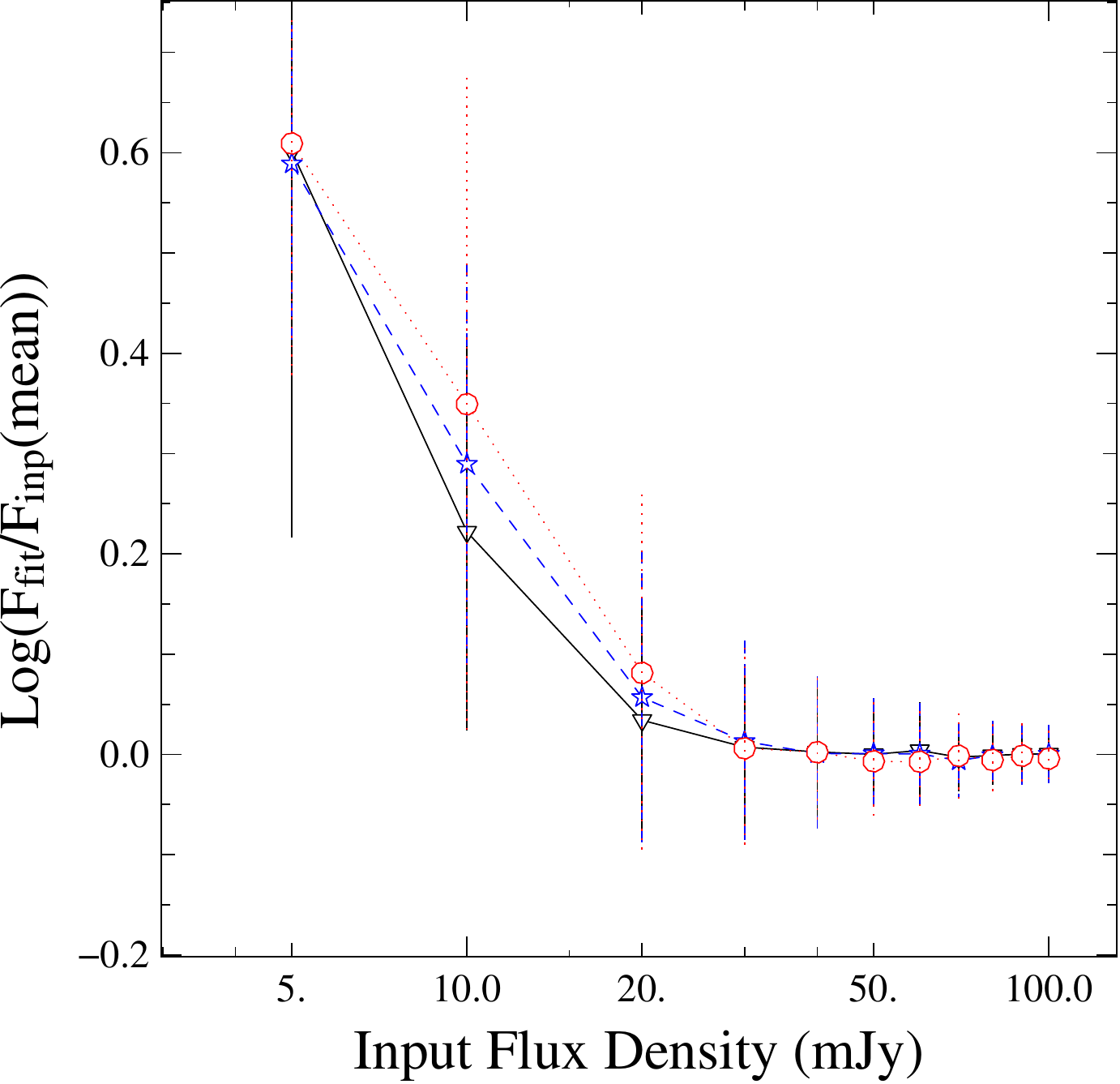}
\end{center}
\caption{Completeness and flux-position accuracy in the catalog at SPIRE wavelengths. In each panel, the black triangles with solid line, blue stars with dash line, and red circle with dot line
represent the results at 250, 350, and 500 $\mu$m. Left panel: Ratio between the numbers of objects detected and the number of sources injected. Middle panel: Radial offset between the sources
injected and the source recovered as a function of the input flux density. Right panel: Logarithm of the ratio between the average recovered flux density and the input one, as a function of the
injected fluxes.}
\label{randomsource_all}
\end{figure*}
%----------------------------------------------------------------------------------------------------------

Once verified, the consistency of our results with respect to the chosen field we calculated the completeness and the flux accuracy of our method at all SPIRE wavelengths and for all the fields, as shown in Fig. \ref{randomsource_all}. The statistics for the tests done at 250, 350, and 500\,\micron\
are also given in Table \ref{V1spire}. The completeness at all SPIRE bands is around 95\% for flux densities of 40 mJy, but the completeness decreases at longer wavelengths at lower fluxes. At 30 mJy, we have a completeness of 88\%, 84\%, and 82\% at 250, 350, and 500 \micron, while the completeness at 20 mJy is reduced to 72\%, 61\%, and 57\% at 250, 350, and 500 \micron. The errors in position are also quite similar, and the differences in absolute value are due to the different pixel sizes, 6\arcsec, 8\arcsec, and 12\arcsec\ at 250, 350, and 500 $\mu$m, respectively (middle panel of Fig.\ref{randomsource_all}).

\subsection{Final catalog}

The final catalogs consist of 52020, 42278, and 18691 unique sources with fluxes above 20 mJy selected at 250, 350, and 500 \micron\ over an area of $\sim$ 55\,deg$^2$. The machine-readable version of the catalog is available at {\it http://www.hevics.org/}. In the overlapping regions of the different fields, the fluxes have been averaged, and the uncertainties have been added in quadrature.

\begin{figure} 
\begin{center} 
\includegraphics[clip=,width=.4\textwidth]{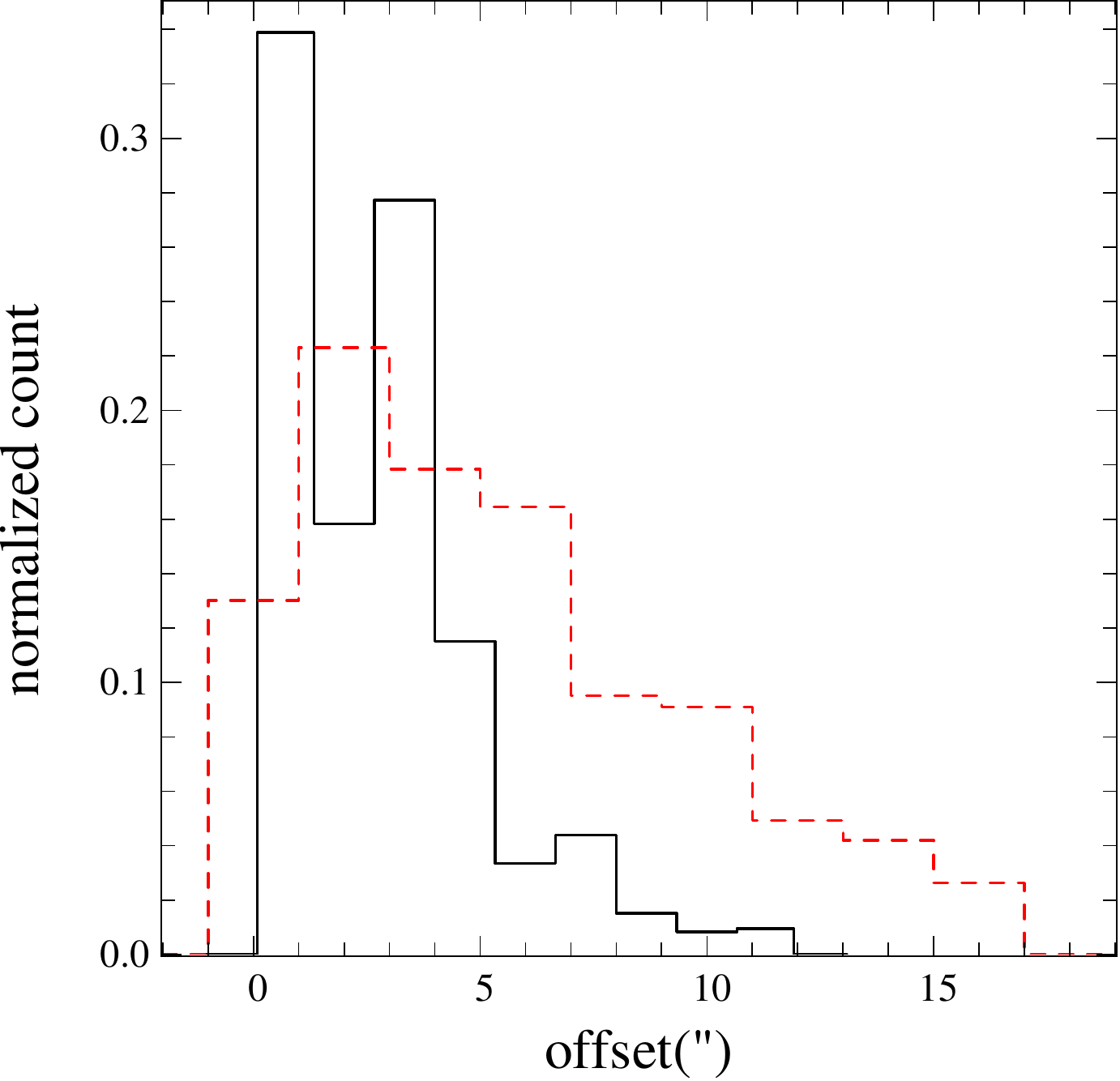} 
\end{center} 
\caption{Normalized histograms of the offset between sources at 250 \micron\ and the associated counterparts at 350 (black solid line) and 500 \micron (red dashed line).}
\label{offset} 
\end{figure}

We considered counterparts to the 250\,\micron\ sources at 350\,\micron\ and 500\,\micron, and their fluxes and positions were determined by positional coincidence with the 250\,\micron\ sources; positional offsets were required to be within a radius equal to half a beam, which are at 12$''$ and 18$''$ at 350\,\micron\ and 500\,\micron, respectively. As shown in Fig. \ref{offset}, at 350 and 500 \micron\ about 90\% and 75\% of the sources have an offset from the 250 \micron\ counterpart below 6\arcsec and 9\arcsec, respectively. These offsets are smaller than the size of a pixel (8\arcsec at 350, and 12\arcsec at 500 \micron) indicating that the number of spurious objects can be considered negligible. The on line version of the catalogs at 350 and 500 \micron\ has also a flag that indicates if there is an object with flux above 20 mJy within a radius of 12\arcsec and 18\arcsec\ at 350\,\micron\ and 500\,\micron\ at 250\,\micron, respectively. About 70\% at 350\,\micron\ and 65\% at 500\,\micron\ of the sources have a counterpart at 250\,\micron, and within these, 68\% and 46\% lie at a radial distance of 9\arcsec (about half beam at 250 \micron).

%----------------------------------------------------------------------------------------------------------
%----------------------------------------------------------------------------------------------------------

%-----------------------------------------------------------------------------------------------------------
\begin{table}\caption[]{Test of completeness and flux accuracy in the HeViCS fields. Column 1: wavelength; column 2: input flux in mJy; column 3, and 4: percentage of sources detected and not
detected;
column 5: percentage of statistical outliers; column 6, 7, and 8: average, median, and standard deviation of the flux densities recovered excluding statistical outliers, respectively.}\label{V1spire}
\begin{center}
\begin{tabular}{ccccc|cccccccccccccc}
\hline
 Wlg & F$_{inp}$ & N$_{det}$& N$_{ndet}$& N$_{out}$ & $F_{avg}$ & $F_{med}$ & $F_{err}$ \\
  \micron & mJy & \% & \% & \% & mJy & mJy & mJy \\
   (1) &  (2) & (3) & (4) & (5) & (6) & (7) & (8)\\ 
&&&& \\ 
 250  &&&&\\
 &      100 & 99.8 & 0.2 & 1.4 & 100 &  99 &   6 \\ 
 &       90 & 99.9 & 0.1 & 2.1 &  90 &  89 &   5 \\ 
 &       80 & 99.5 & 0.5 & 1.9 &  80 &  79 &   6 \\ 
 &       70 & 99.8 & 0.2 & 1.6 &  70 &  69 &   6 \\ 
 &       60 & 98.9 & 1.1 & 1.7 &  60 &  59 &   6 \\ 
 &       50 & 98.0 & 2.1 & 2.1 &  50 &  49 &   5 \\ 
 &       40 & 95.4 & 4.6 & 1.8 &  40 &  40 &   6 \\ 
 &       30 & 88.5 & 11.5 & 1.6 &  30 &  30 &   6 \\ 
 &       20 & 74.2 & 25.8 & 1.5 &  21 &  21 &   5 \\ 
 &       10 & 36.4 & 63.5 & 0.8 &  17 &  15 &  10 \\ 
 &        5 & 15.0 & 85.0 & 0.4 &  21 &  16 &  19 \\ 
  350  &&&&\\
 &      100 & 99.9 & 0.1 & 1.7 & 100 &  99 &   6 \\ 
 &       90 & 99.9 & 0.1 & 1.6 &  90 &  89 &   6 \\ 
 &       80 & 99.7 & 0.3 & 1.7 &  80 &  79 &   6 \\ 
 &       70 & 99.5 & 0.5 & 1.9 &  70 &  69 &   6 \\ 
 &       60 & 98.9 & 1.1 & 1.5 &  60 &  59 &   6 \\ 
 &       50 & 97.3 & 2.7 & 1.5 &  50 &  49 &   6 \\ 
 &       40 & 93.0 & 7.0 & 1.9 &  40 &  39 &   6 \\ 
 &       30 & 84.2 & 15.8 & 1.6 &  31 &  30 &   6 \\ 
 &       20 & 61.5 & 38.5 & 1.1 &  23 &  22 &   7 \\ 
 &       10 & 25.1 & 74.9 & 0.5 &  18 &  17 &   8 \\ 
 &        5 & 11.7 & 88.3 & 0.5 &  20 &  18 &  10 \\ 
  500  &&&&\\
 &      100 & 100.0 & 0.0 & 1.3 &  99 &  99 &   6 \\ 
 &       90 & 100.0 & 0.0 & 0.9 &  90 &  89 &   6 \\ 
 &       80 & 99.9 & 0.1 & 1.6 &  79 &  79 &   6 \\ 
 &       70 & 99.6 & 0.4 & 1.3 &  70 &  69 &   6 \\ 
 &       60 & 99.3 & 0.7 & 1.1 &  60 &  59 &   6 \\ 
 &       50 & 97.5 & 2.5 & 1.5 &  50 &  49 &   6 \\ 
 &       40 & 94.1 & 5.9 & 1.3 &  40 &  40 &   6 \\ 
 &       30 & 82.1 & 17.9 & 1.2 &  31 &  30 &   6 \\ 
 &       20 & 57.9 & 42.1 & 0.7 &  23 &  23 &   7 \\ 
 &       10 & 24.1 & 75.9 & 0.5 &  21 &  19 &  12 \\ 
 &        5 & 10.5 & 89.5 & 0.5 &  21 &  18 &  11 \\ 
 \hline
\end{tabular}
\end{center}
\end{table}
%-----------------------------------------------------------------------------------------------------------

\section{Analysis}
\label{analisi}
%----------------------
\subsection{Number counts}

\begin{figure*}
\begin{center}
\includegraphics[clip=,width= .33\textwidth]{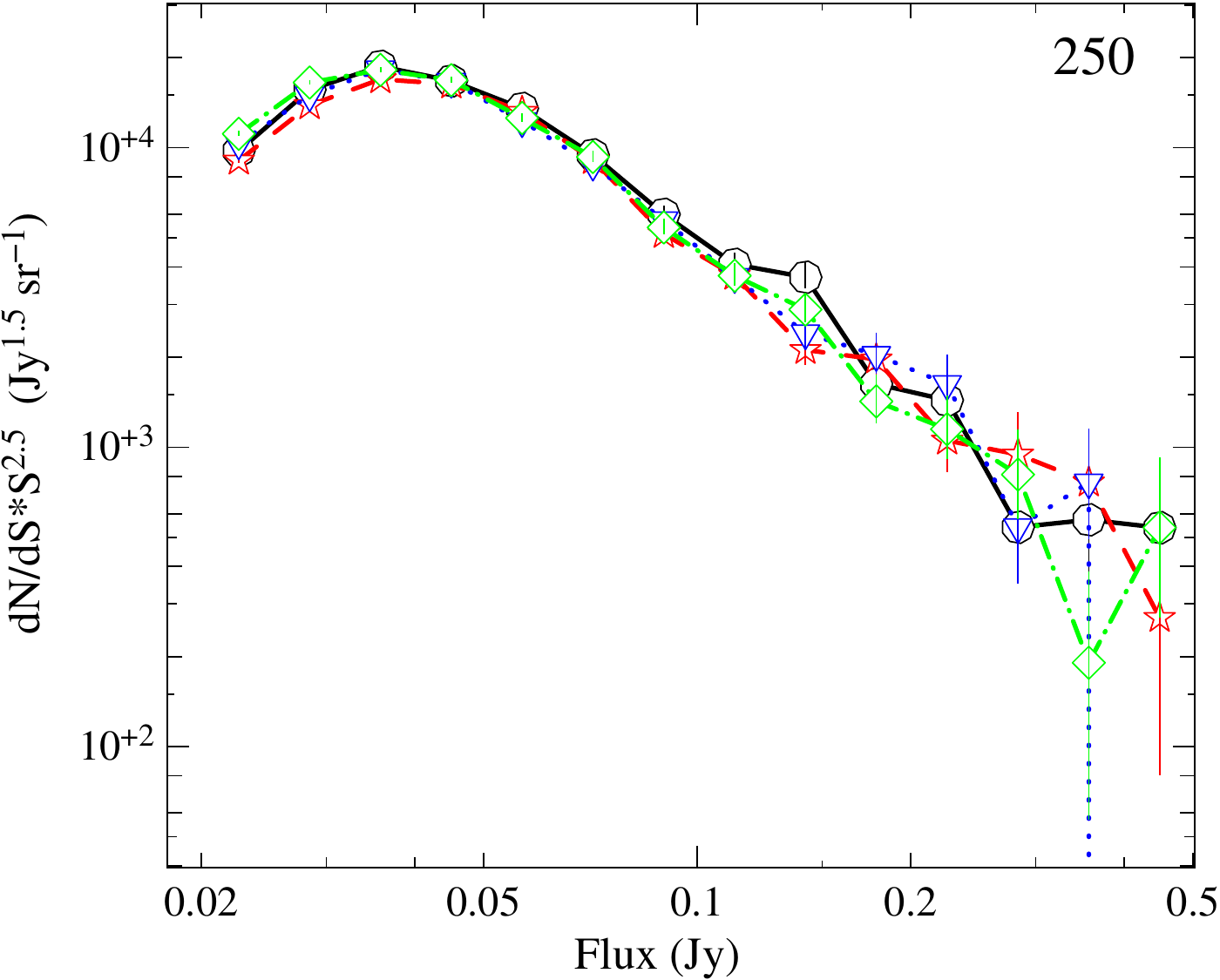}
\includegraphics[clip=,width= .33\textwidth]{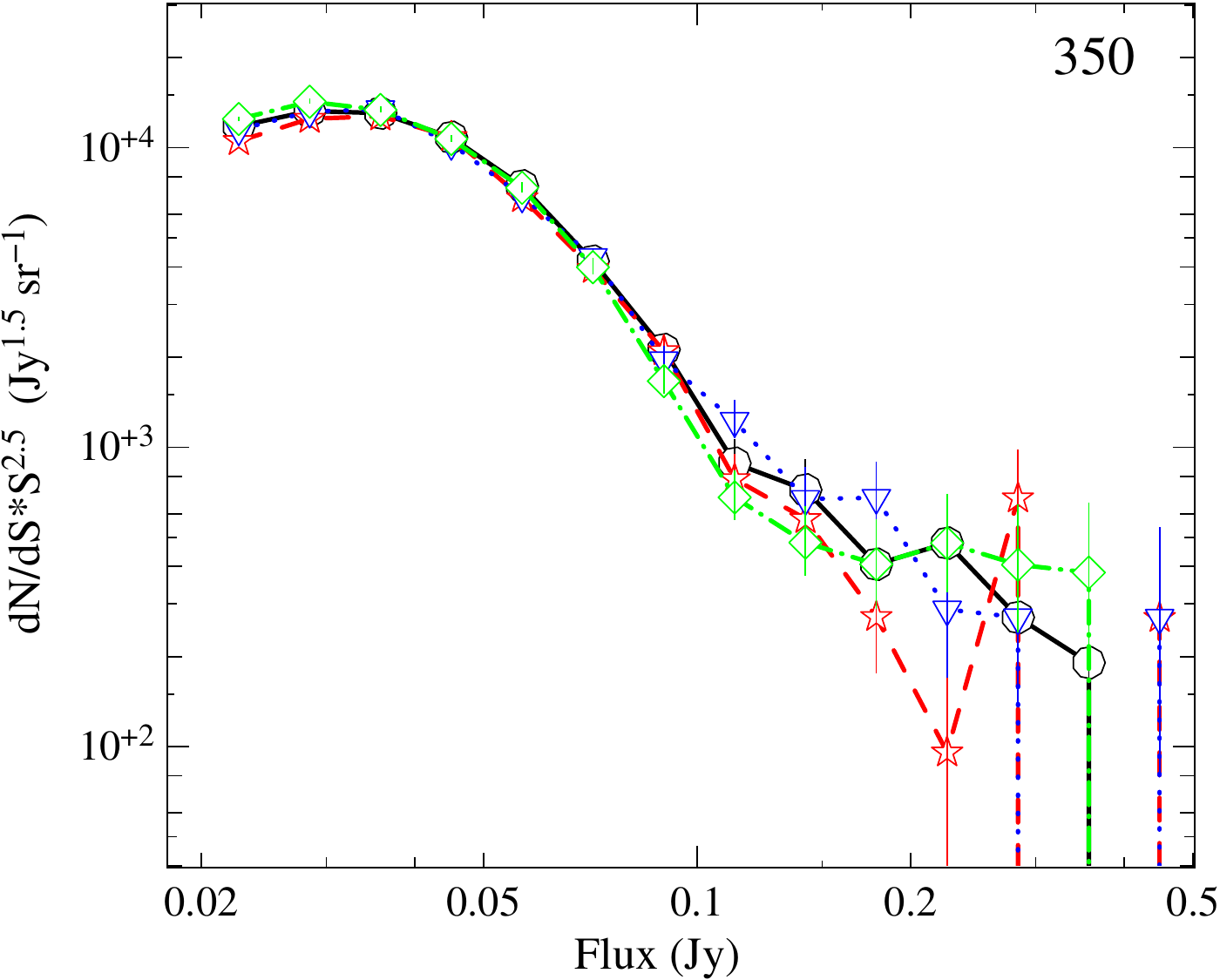}
\includegraphics[clip=,width= .33\textwidth]{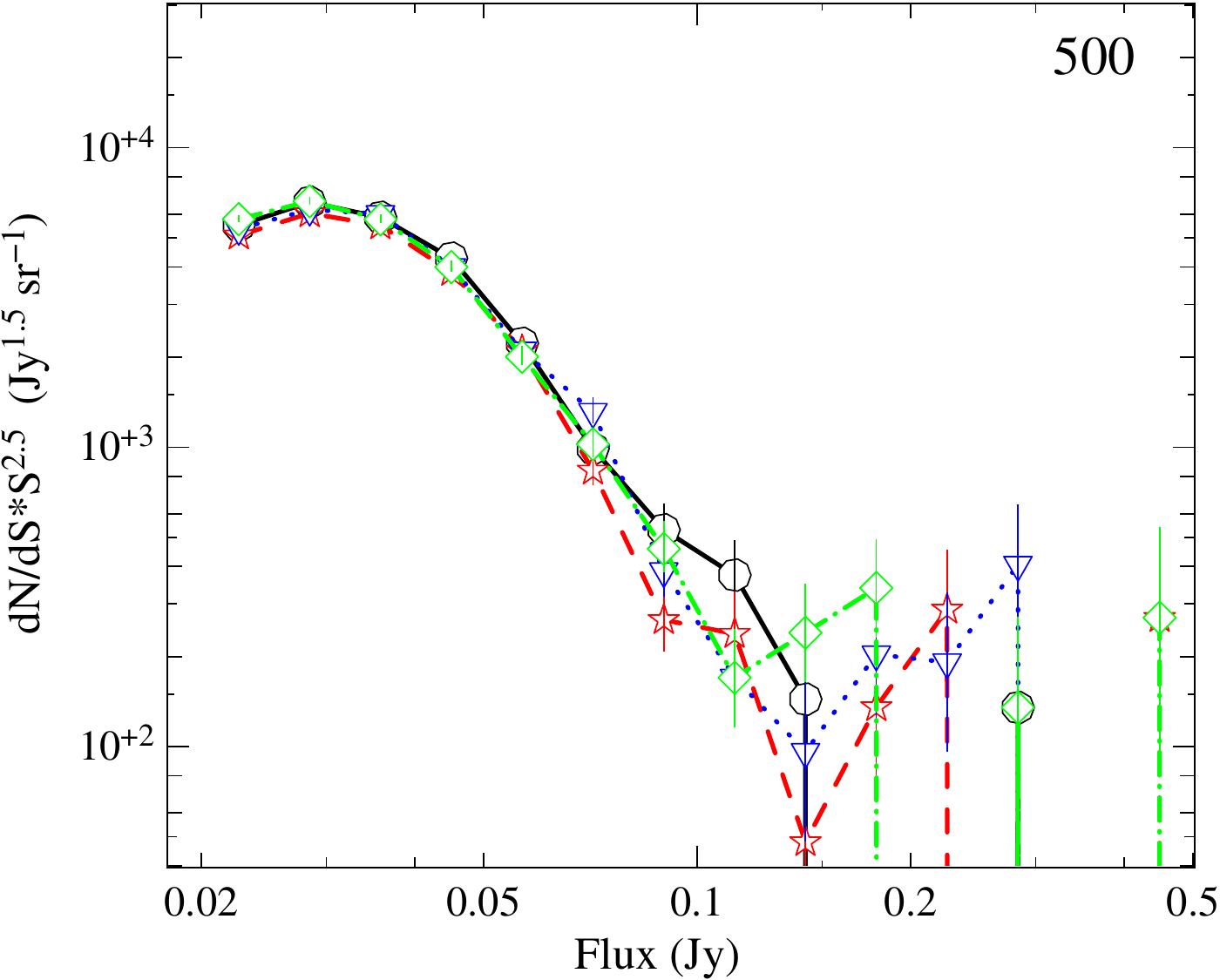}
\end{center}
\caption{Number counts estimated in each field at 250 (left), 350 (middle), and 500 (right) $\mu$m. Green diamonds, black circles, blue triangles, and red stars indicate V1, V2, V3, and V4 field,
respectively.}
\label{ncou}
\end{figure*}

As a preliminary test, we analyze the number counts for each different field separately to
investigate the reliability of the procedures used to extract sources and to
perform the photometry. Since the effect of clustering due to the background sources can be
considered negligible in our maps because of the large map size, we expect not to find
significant variations among different fields, at least, if a significant fraction of the point sources are not within the cluster. This is confirmed in Fig.
\ref{ncou}, which shows the number counts of the
four HeViCS fields, separately. The units in the ordinate are used to have a direct representation of the evolution in luminosity of the sources, since the number counts in an isotropic Euclidean Universe with no evolution would
appear as a straight horizontal line \citep{lon}. There is no relevant differences between the fields with a high number of extended sources (in particular the V2 field) and the one in which this component is negligible. This consistency is verified at
each flux value, in particular for faint fluxes, indicating that Virgo cluster galaxies give a negligible contribution to the total number counts. This also suggests that the Virgo cluster does not contain an infrared-bright, optically-faint population of galaxies. There is some difference between different fields for sources at fluxes above 150\,mJy, but these differences are mainly driven by the relatively small number of sources at those fluxes.

 %-----------------------------------------------------------------------------------------------------------
%-----------------------------------------------------------------------------------------------------------

\begin{figure*}
\begin{center} 
\includegraphics[clip=,width= .33\textwidth]{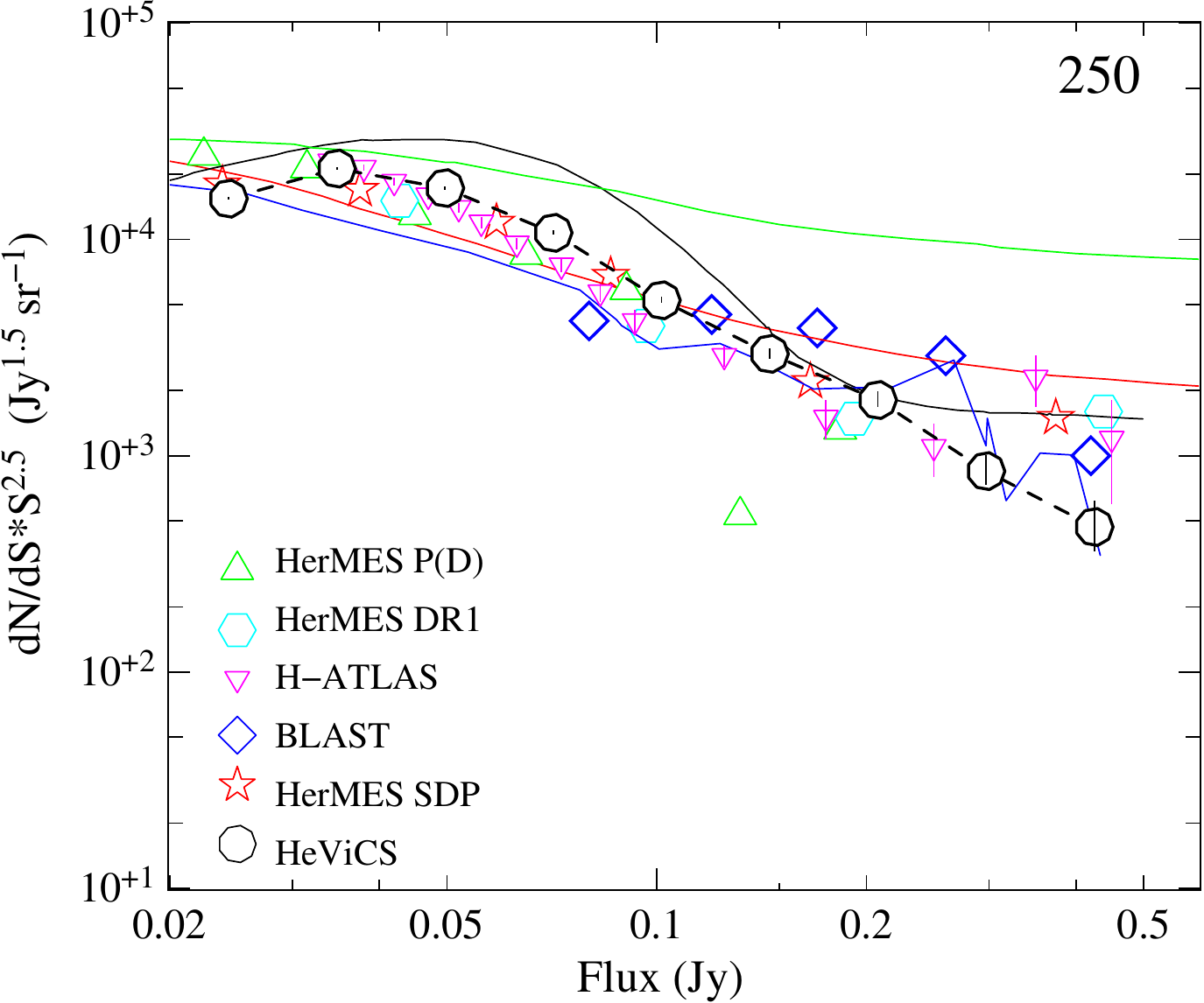} 
\includegraphics[clip=,width= .33\textwidth]{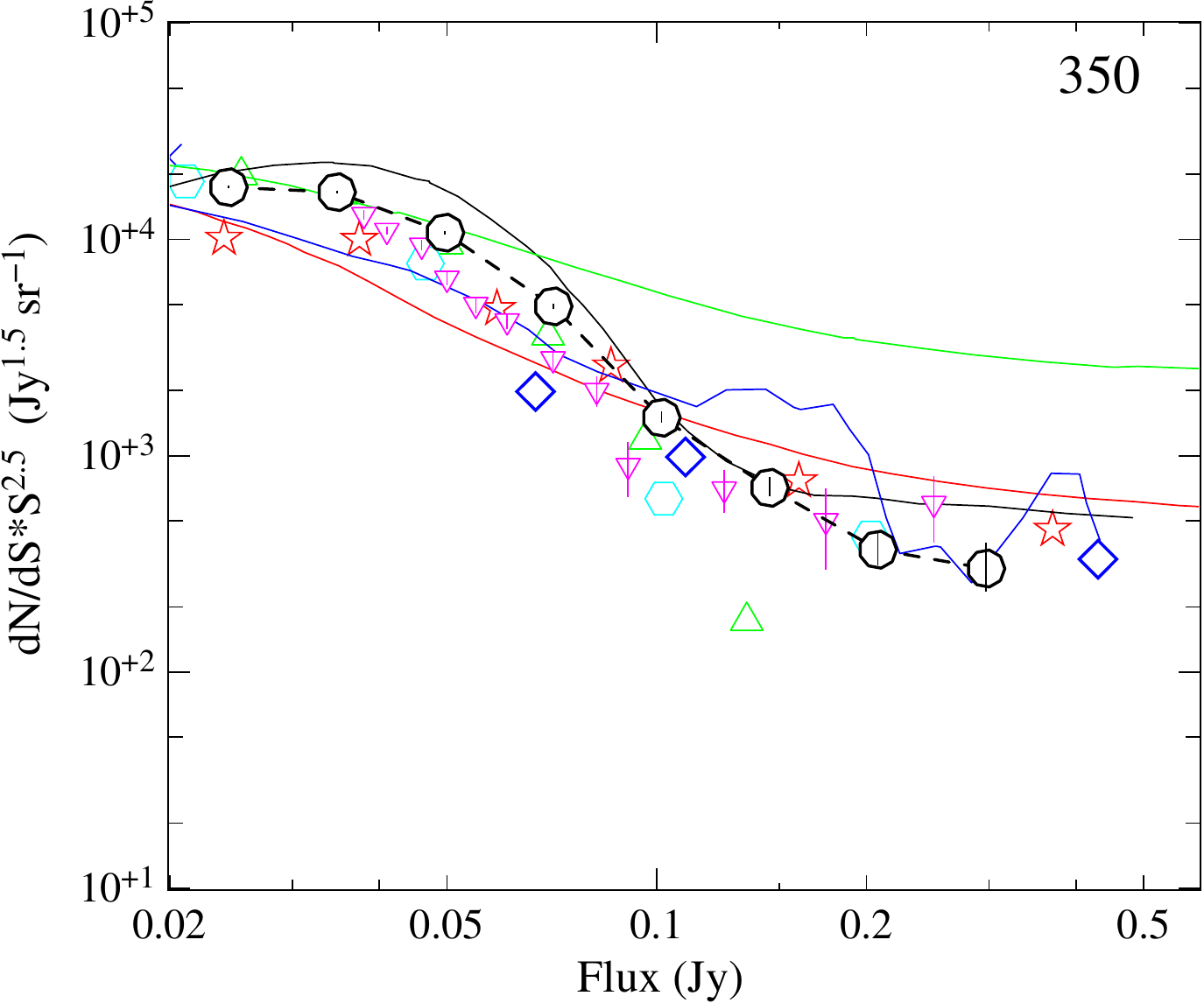} 
\includegraphics[clip=,width= .33\textwidth]{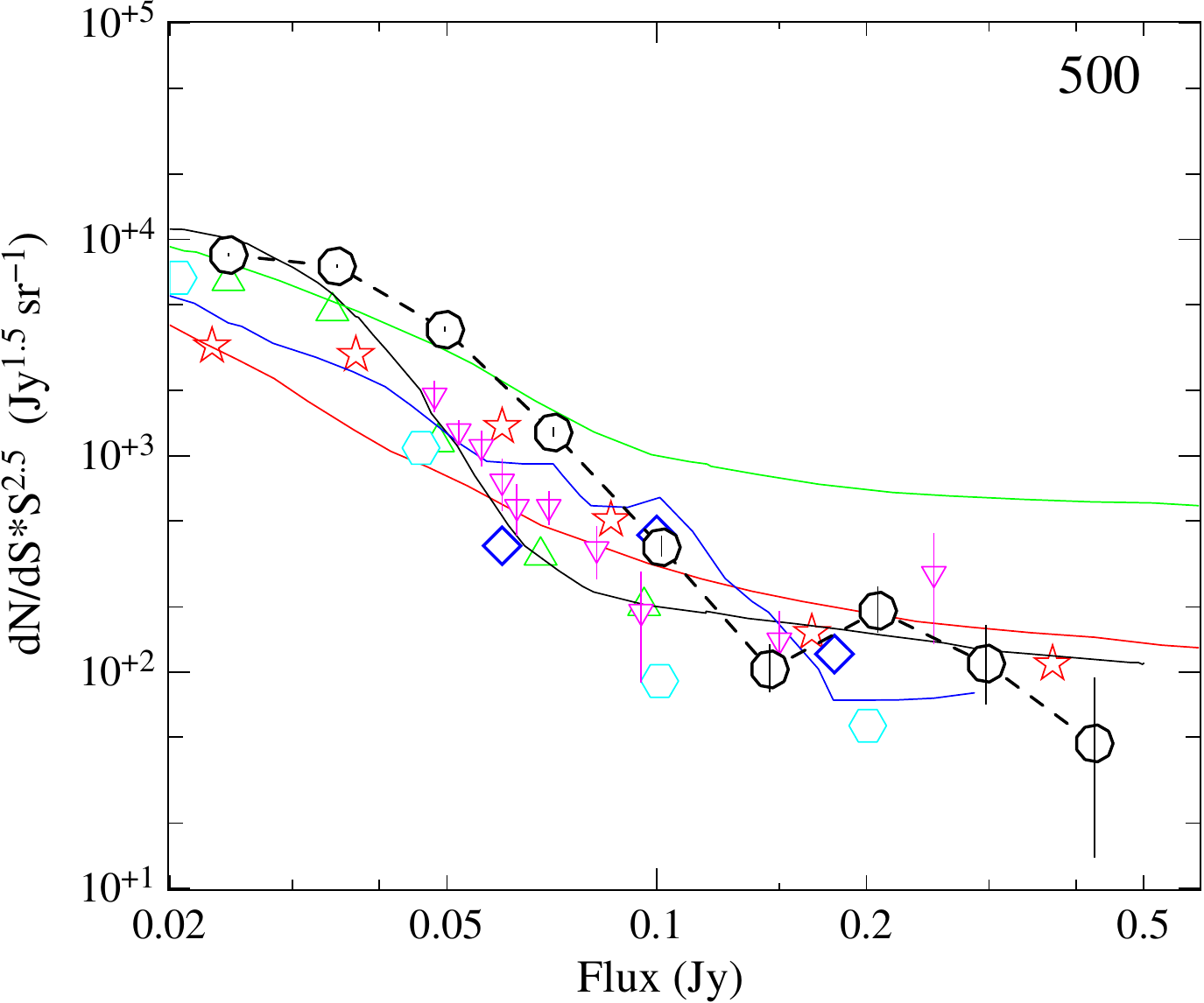} 
\end{center}
\caption{Number counts
(black circles) at 250 (left), 350 (middle), and 500 (right) $\mu$m. Magenta
triangles, red stars, blue diamonds, green triangles, and cyan hexagons show
the observations from H-ATLAS \citep{cle}, HerMES SDP \citep{oli}, BLAST
\citep{pat}, HerMES DR1 \citep{bet}, and the results of a $P(D)$ (probability
of deflection) analysis applied to the HerMES SDP data from \cite{gle}, respectively. The green, black, blue, and red solid lines show the models of \citet{lac2},
\citet{neg}, \citet{val}, and \citet{bet2}.}
\label{ncou2}
\end{figure*}

The HeViCS number counts for the combined fields corrected for the incompleteness are shown in
Fig. \ref{ncou2} with the result of other analyses: H-ATLAS \citep[{\it
Herschel } Astrophysical TeraHertz Large Area Survey,
][]{cle}, HerMES ({\it Herschel} Multi-Tiered Extragalactic Survey) SDP
\citep[][]{oli} and DR1 \citep[][]{bet}, together with
a statistical analysis from \cite{gle}. Finally,
we also include the results of BLAST \citep[Baloon-borne
Large-Aperture Submillimeter Telescope][]{pas}, an
experiment realized using a precursor of the SPIRE instrument. 

A characteristic
feature at all the wavelengths is an increase in the number counts at $F\la$ 200\,mJy at
250, 350, and 500 $\mu$m, indicating a strong evolution for the galaxy
populations at faint fluxes. Comparison of the number counts at different wavelengths shows that the slope
of the faint end where the statistics are good (40 $\la F \la$150\,mJy) steepens going from 250\,\micron\ to the longer wavelengths.
Below this flux level, the number counts flatten and decrease because of
the confusion. 

There is quite good agreement between our data and the
results of other {\it Herschel} surveys at faint fluxes below 200 and 180\,mJy at 250 and 350 $\mu$m, respectively. However at 500 \micron, the number of sources at low flux densities is higher
with respect to other surveys, which is probably an effect of our source extraction method. At flux densities below 30\,mJy, we found $\sim$ 90\% more sources with respect to the HerMES SDP survey; this number decreases to $\sim$ 40\% when we consider sources with a flux above 60 mJy. A similar tendency is seen comparing our results to the HerMES DR1 \citep{bet} data and the deeper HerMES statistical analysis done by \cite{gle},
which resolved 43\% of the far-infrared background at 500\,\micron. The subtraction of the sources from the timeline allowed us to recover more objects with respect to other methods, and this effect is
more relevant when the beam is larger, because the probability in this case that sources are blended together, or partially covered by a brighter source is higher with respect to smaller beamsizes.
The increased slope of the number counts at wavelengths above 350 $\mu$m might be due to the presence of
proto-spheroidal galaxies, which are mostly evolving passively at lower
redshift, and became more active at redshift $z \ge$ 1.5. 
Because of the strongly negative {\it k}-correction at sub-mm wavelengths, 
the number of these galaxies is
weakly dependent on the redshift, resulting in a steep rise of the
number counts at $\lambda \ge$ 350 $\mu$m. 

Together with the results of other analyses, we compare the HeviCS number of counts with other theoretical models. There are two main approaches to theoretically understanding the number counts of galaxies. A first method, `forward evolution', is based on semi-analytic models of galaxy formation and evolution. In these models, the time evolution of the observational properties of galaxies are predicted using prescriptions for the physics of the baryonic components that are embedded in the large structure skeleton of the dark matter, following the standard $\Lambda$CDM cosmology. An example of this method is shown in \cite{lac}, which has been applied with success to {\it Spitzer} data, and extended to {\it Herschel} FIR
wavelengths \citep{lac2}.

The second method, `backward evolution' \citep[e.g.][]{bet3}, starts from the observed luminosity and density
function at $z = 0$ and then assumes a parametric form for the luminosity evolution. A set of spectral energy distribution (SED)
templates representative of different galaxy populations are made to evolve in density and luminosity according to the luminosity function parameterization. This approach can give
different results according to the different choices of the galaxy SED
templates, and the different physical processes that are taken into account in
the modeling. We choose the model of \cite{val} for our comparison with the data, which includes an evolution with the redshift for the AGN contribution to the
total FIR luminosity, and the model of \cite{neg}, which includes the
contribution of gravitational lensing to the observed fluxes, a contribution
not negligible at wavelengths $\ga$ 500 $\mu$m \citep[e.g.][]{neg2}. 

Analyzing GOODS-{\it Herschel} data, \cite{elb} found that the evolution of IR
galaxies follows two distinct star-formation mechanisms. Most galaxies
follow a main sequence with a tight correlation between stellar mass and 
star formation, independent of luminosity and redshift. A smaller galaxy
population, less than 20\%, shows a starburst phase, with high specific star-formation rate \citep[e.g.,][]{magnelli12}. Supported by these results, \cite{bet2} proposed a model that
considers the AGN contribution, gravitational lensing, and assumed only two types
of galaxy SEDs for the main sequence and the starburst population. They also introduce
an evolution with redshift for the SED templates, following the recent results
of \cite{mag}, which showed that the mean radiation field of a galaxies is
stronger at higher redshift. There is a substantial amount of observational evidence that suggests a redshift evolution of the total IR luminosity and dust temperature (e.g. \citealt{amb}, \citealt{elb2}, \citealt{hwa}, \citealt{cas}, and \citealt{mag14}).  

We compare our observed number counts with
the predicted number counts at each SPIRE wavelength in Fig.
\ref{ncou2}, according to the models of \citet{lac2,neg,val} and \citet{bet2}.
The \citet{lac2} model predictions tend to exceed the observed ones
at all fluxes. This could be because the model employs a top-heavy stellar initial mass function \citep[similarly
to][]{bau} that produces more dust per unit star-formation rate. In
this way, they match the bright end of galaxy counts at 850 $\mu$m, but the number counts are systematically overestimated at lower
fluxes. 
The model of \cite{val} gives the best agreement with our data
\citep[see also][]{gle}, indicating the
importance of an AGN contribution in shaping the galaxy SEDs, although the source
fluxes $\ga$ 100\,mJy at 350 and 500 $\mu$m tend to be overestimated. The models by \cite{neg}
reproduce quite well the slope and the evolution at lower fluxes but
overestimate the number counts at 250 and 350 $\mu$m. At 500 $\mu$m, the comparison with our data is good, but the increase in the number of counts occurs at lower fluxes with respect to our data.

Generally, the models tend to overpredict the observed counts at bright flux and underpredict them at faint ones.
These discrepancies between model predictions and observations
can arise from several factors. For example, the
evolution of the SEDs with the redshift is poorly known, and the SEDs library
is well calibrated only in the local Universe, where the $k$-correction is
small. Another effect that can play a role is the AGN contribution to the dust
heating \citep[e.g.][]{alm,gra,far,saj}, although recent results suggest that dust emission due to heating by stars dominates that from the heating by AGNs at wavelengths larger than $\sim$ 30 \micron\ \citep{hat}. The {\it Chandra} X-ray images of luminous
BzK galaxies at $z \sim$ 2 \citep{dad} exhibit soft-to-hard X-ray ratio
typical of heavily obscured AGN, but deeper observations have shown that only
25\% of luminous BzK galaxies host heavily obscured AGN \citep{ale} and the
rest are star-forming galaxies with unobscured AGN.  In any case, the
difficulties in reproducing the number counts at fluxes $\ga$ 100\,mJy
underline the importance of the Rayleigh-Jeans side of the SED to characterize
the evolution of FIR galaxy population, a region for which {\it Herschel}
is fundamental. 

%-----------------------------------------------------------------------------------------------------------
%-----------------------------------------------------------------------------------------------------------

\subsubsection{Contribution to the number counts from different redshift bins}

We searched for the optical counterparts of the HeViCS FIR sources in the $r$-band of the SDSS DR10 survey \citep{ahn} using the likelihood ratio method described in \cite{sut}. This method associates a parameter, the reliability (${\mathcal R}$), to each optical source, that quantifies the probability that the SDSS source considered is the true optical counterpart of the FIR object of reference. To determine ${\mathcal R}$, we account for the possibility that the counterpart is below the optical magnitude limit with the probability to find serendipitously a spurious source in a given position. To define the optical matching sample, we extracted a 55\,deg$^2$ cutout from the $r$-band SDSS catalog, perfectly overlapping with the HeViCS surveyed area. This contains 1\,714\,999 sources with a completeness limit of 95\% at $r \approx 22.2$ AB mag in optical imaging \citep{ahn}. The Milky Way extinction corrected SDSS $r$-magnitudes were used to compute the surface density distribution of the optical objects.

\begin{figure}\begin{center}
\includegraphics[clip=,width= .35\textwidth,angle=90]{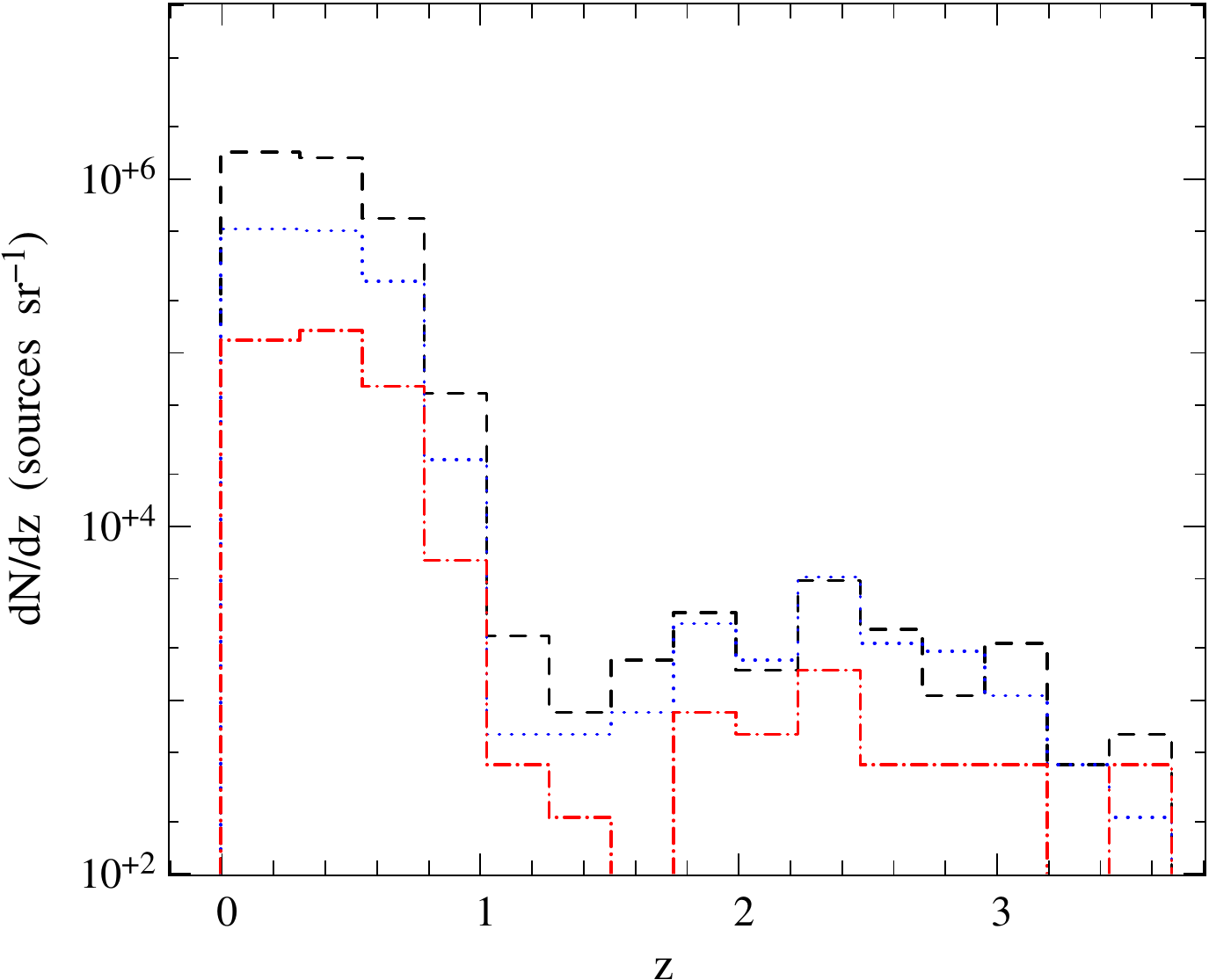}
\end{center}
\caption{Redshift histogram for the sources that have an optical counterpart in the SDSS catalog with a reliability above 80\% at 250 (black dashed line), 350 (blue dotted line), and 500 (red dot-dashed line)
$\mu$m. Spectroscopic redshifts have been considered where availables, photometric redshift are considered otherwise.}\label{redshi}
\end{figure}

\begin{figure*}\begin{center}
\includegraphics[clip=,width= .32\textwidth]{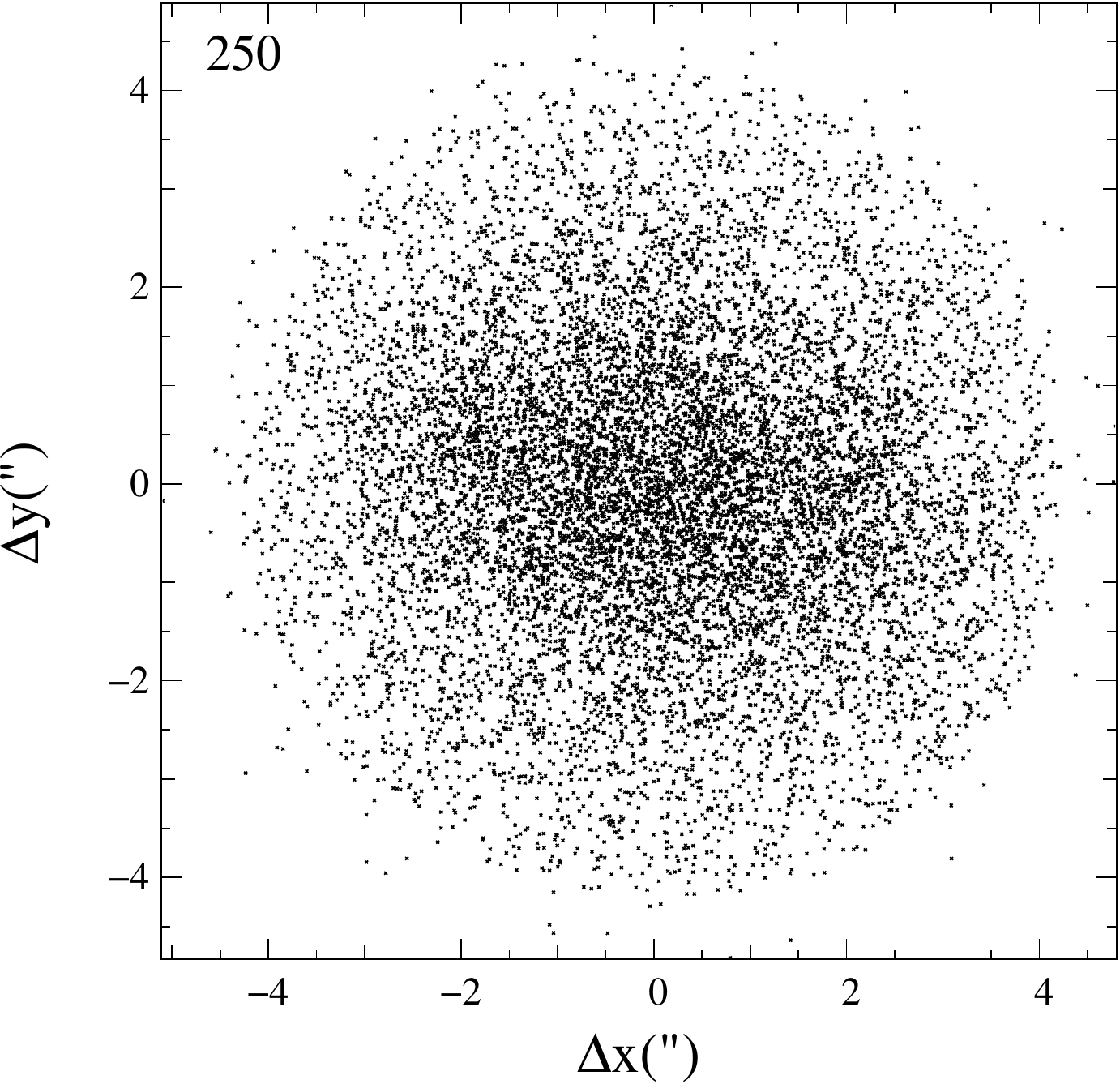}
\includegraphics[clip=,width= .32\textwidth]{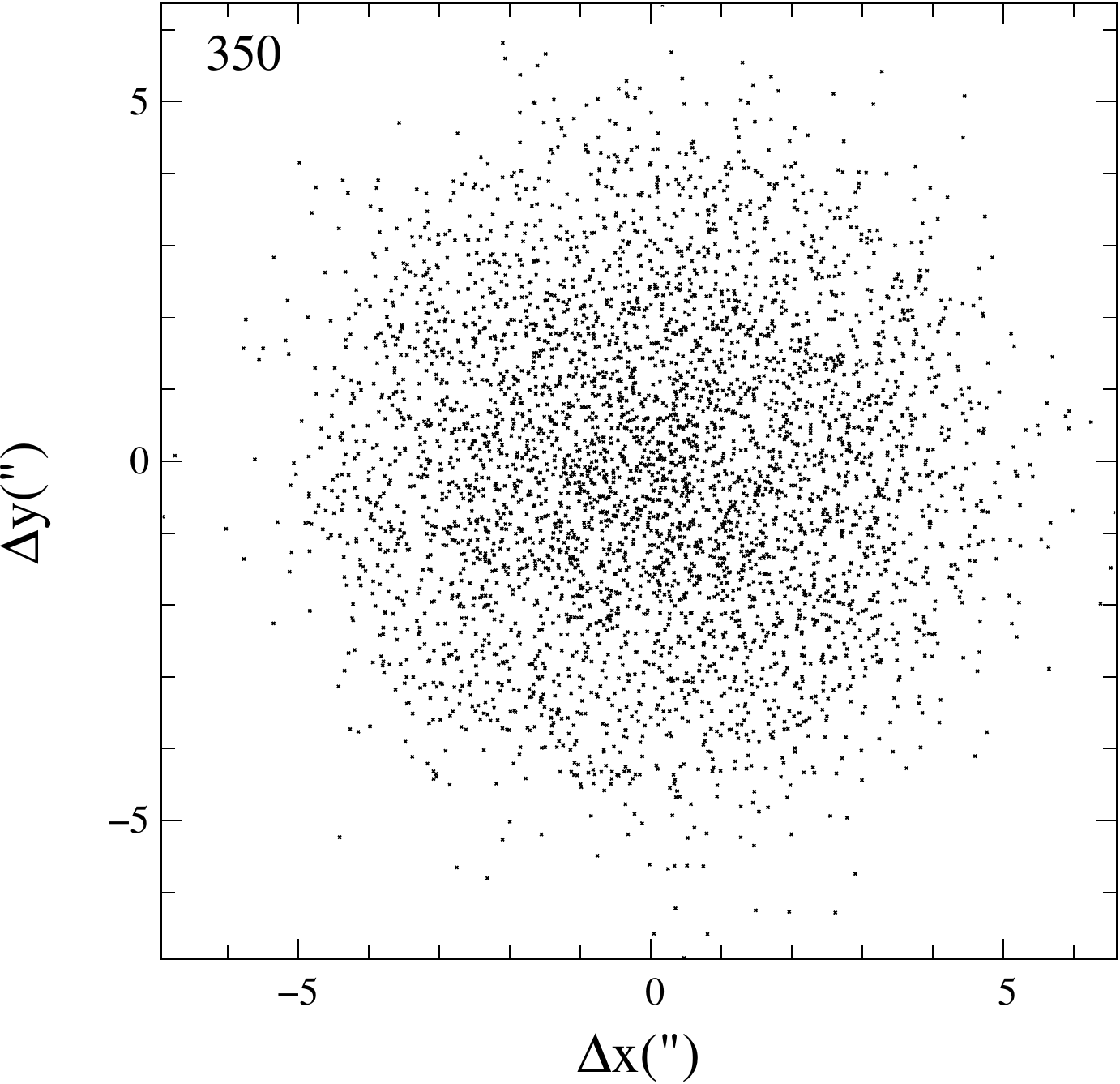}
\includegraphics[clip=,width= .32\textwidth]{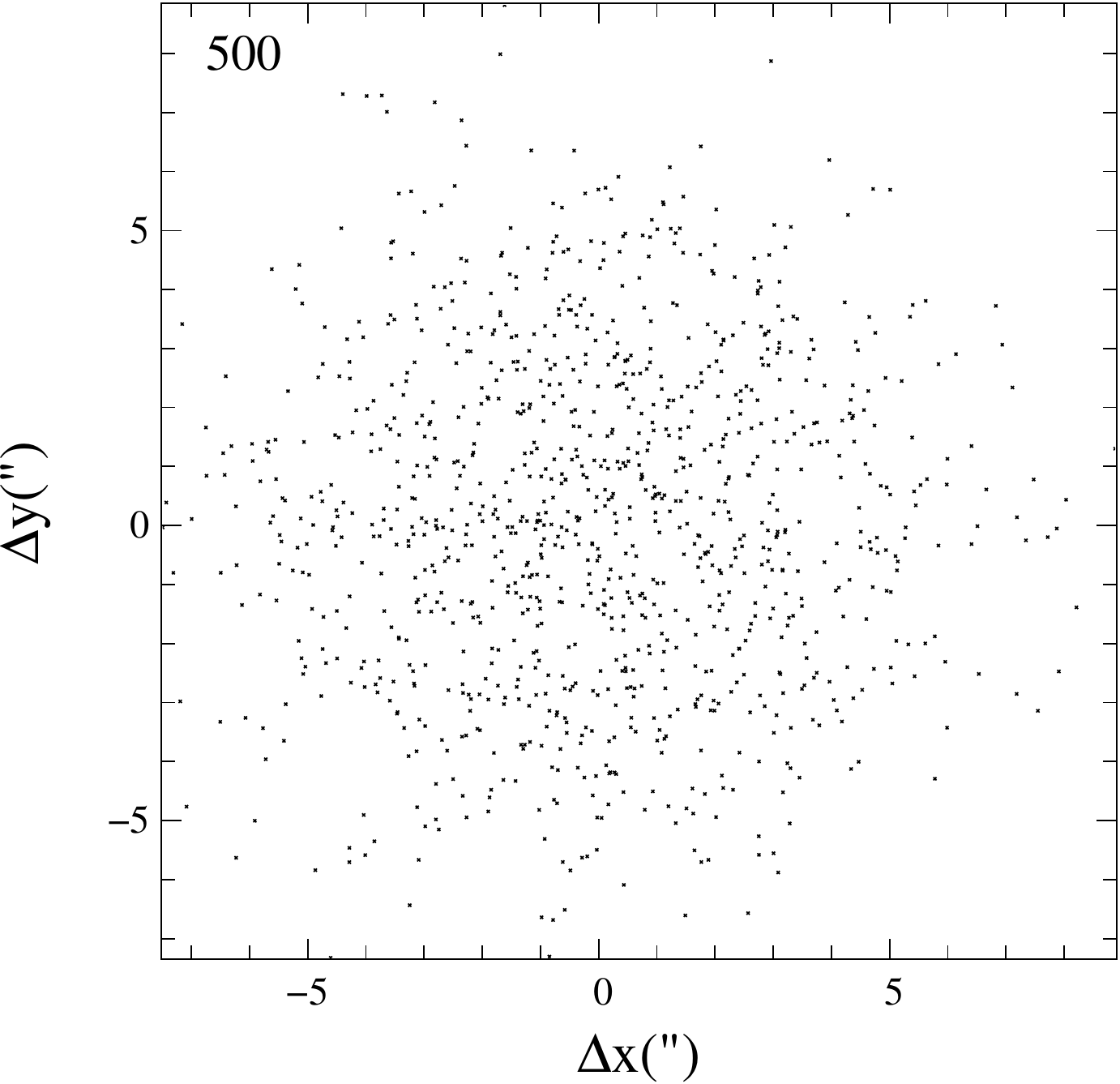}
\end{center}
\caption{Offsets between the HeViCS coordinates and its associate optical counterpart. In the left, middle, and right panels, the offsets at 250, 350, and 500 \micron\ are shown, respectively}\label{offradec}
\end{figure*}

The positional uncertainties in the sky coordinates were derived for each infrared source from the timeline fitting; however, we further considered that error radii less than a few arcsec are
unrealistic given the possibility of systematic offsets \citep[see][for details]{smi}. We thus assumed 2\farcs5, 3\farcs4, and 5\arcsec\ as minimum error radii for the HeViCS 250, 350, and 500\,$\mu$m sources,
respectively. For each HeViCS source, the $r$-band identification with the ${\mathcal R} >$ 80\%  was taken as the real counterpart, corresponding to 40\%, 25\%, and 19\% of the total sources at 250, 350, and 500 \micron, respectively. The average redshift of the catalog having significant detections at all SPIRE wavelengths is $z$ $\sim$ 0.3 $\pm$ 0.22 and 0.25 (16-84 percentile). The relatively large spread arises because $\sim$ 80 \% of the sources are at $z <$ 0.5, while the remaining 20\% are at higher redshift with 0.5 $<$ $z$ $<$ 3.5, as shown in the histograms of Fig. \ref{redshi}. In this figure, we have considered the spectroscopic redshift where available. For the others, we use the photometric redshift obtained with a robust fit on spectroscopically observed objects with similar colors and inclination angle. 
In Fig. \ref{offradec}, we show the offset between the HeViCS source and the associated optical counterparts. Most of the sources are within a radius smaller than 6\arcsec, which is the smallest pixel size of SPIRE instrument, confirming the reliability of this procedure.

%-----------------------------------------------------------------------------------------------------------
%-----------------------------------------------------------------------------------------------------------

\begin{figure*}
\begin{center} 
\includegraphics[clip=,width= .33\textwidth]{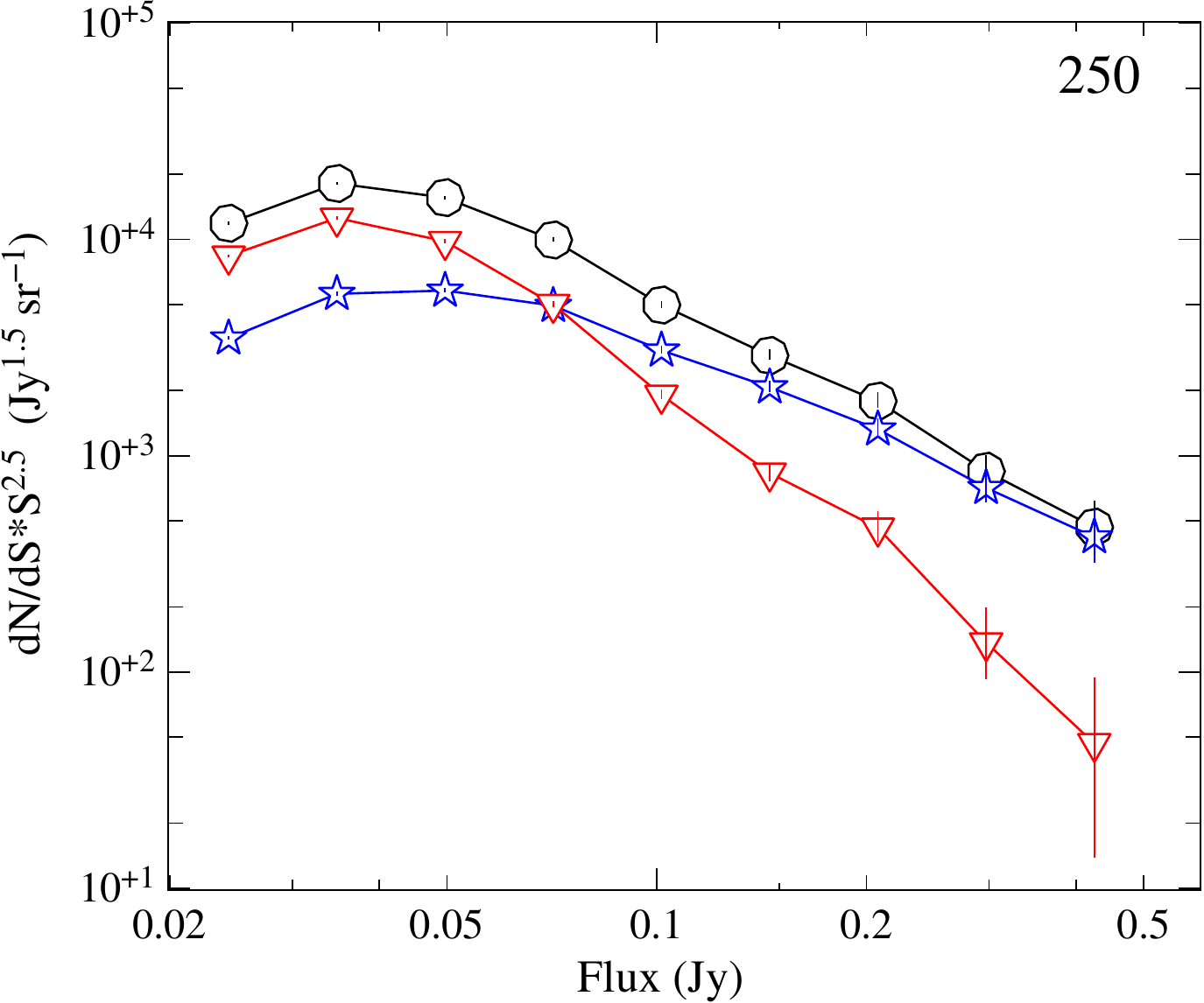} 
\includegraphics[clip=,width= .33\textwidth]{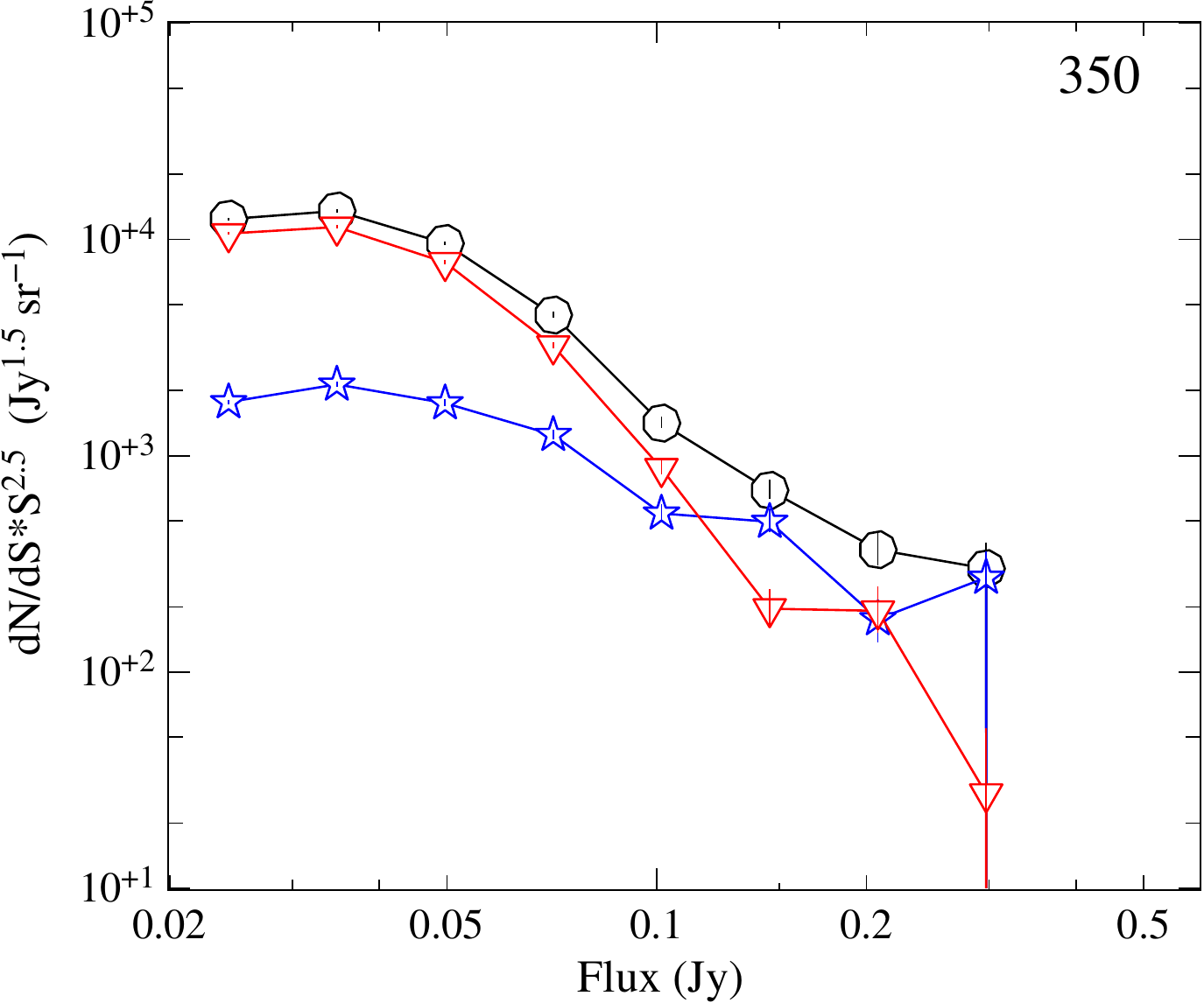} 
\includegraphics[clip=,width= .33\textwidth]{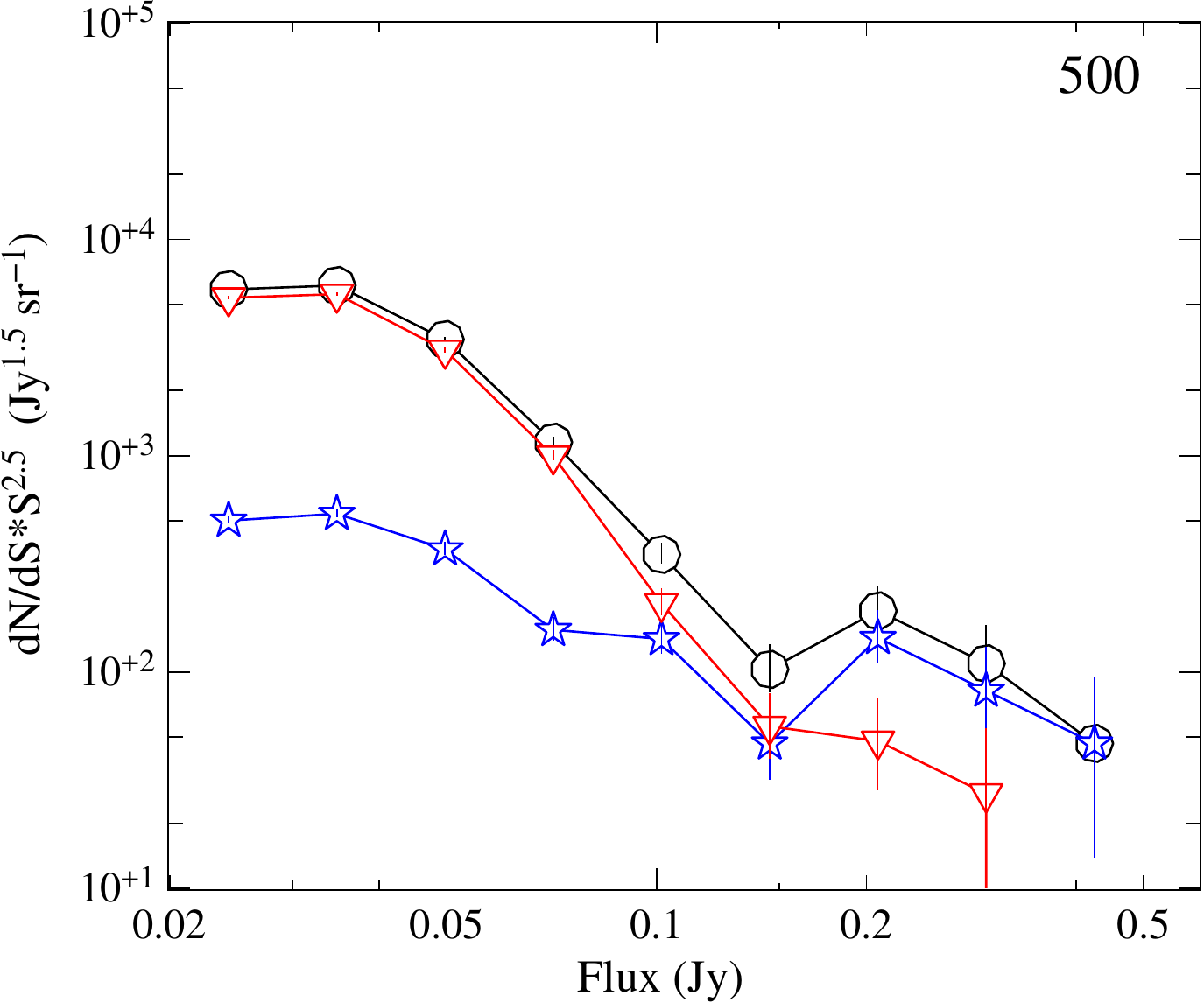} 
\end{center}
\caption{Number counts (black circles) at 250 (left), 350 (middle), and 500 (right) $\mu$m with the component given by the sources with SDSS counterparts (blue stars) and no-SDSS counterpart (red
triangles).}
\label{nc_red}
\end{figure*}

Clearly, the redshift distribution considered for the sources with an optical counterpart in SDSS is biased towards low redshift galaxies. Because the SDSS catalog is 95\% complete for $r=$ 22.2 AB mag, the
cross correlation with our FIR catalogs give us the possibility to estimate the contribution of the low redshift component to the total number counts. In Fig. \ref{nc_red}, we show the number of counts
of the HeViCS sources, the subsample with optical counterparts and redshift estimation, and the component without redshift. Above 100 mJy, the number counts are dominated by the component with the optical counterpart that forms the bulk of the bright end of the distribution, while the component without redshift becomes more important at flux densities below $\sim$ 60 mJy. 

The number counts of objects with no optical counterparts for any SPIRE wavelength are steeper than the sample with optical counterparts, indicating a more rapid density evolution for the former component. Moreover, the contribution from the component with no optical counterparts to the total number counts becomes more relevant for an increasing wavelength. This is because the peak of IR SED is around 200 $\mu$m, and by increasing the wavelength, we consider sources statistically at higher redshift, where the SDSS is less efficient in detecting sources. At fluxes above 100 mJy, the number counts at all SPIRE wavelengths, as expected, is dominated by the component with optical counterpart, which are on average star-forming galaxies at $z \sim$ 0.3.

%-----------------------------------------------------------------------------------------------------------
%-----------------------------------------------------------------------------------------------------------

%-----------------------------------------------------------------------------------------------------------
%-----------------------------------------------------------------------------------------------------------

\subsection{Far-infrared colors}

\begin{figure*}
\begin{center}
\includegraphics[clip=,width= .99\textwidth]{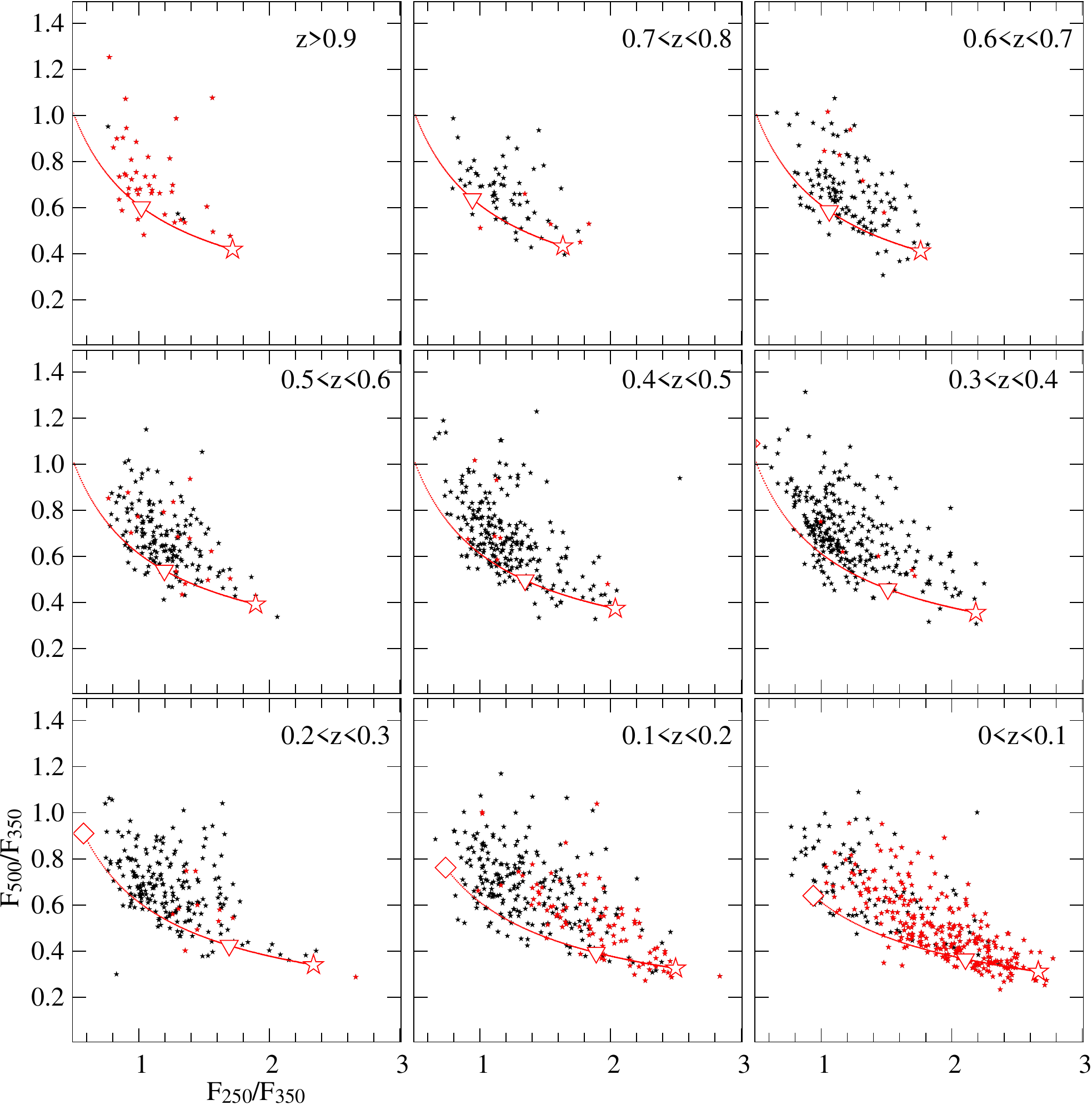} 
\end{center}
\caption{Position in a color-color diagram at different redshift bins of the HeViCS sources with photometric (black stars), spectroscopic (red stars) redshift, and reliable optical counterparts. The
red solid line shows a reference curve for a modified black body with $\beta$ = 2, 10 $\le T \le$ 30, and the redshift equals to the average of the bin. Rhombus, triangles, and stars indicate a modified
black body with $\beta$ = 2 and temperature of 10, 20, and 30 K, respectively.}\label{redshi_evol}
\end{figure*}

In this section, we investigate the possibility, as seen, for example, in H-ATLAS \citep{amb}, to use the color-color diagram at SPIRE wavelengths to statistically constrain the redshift distribution of
the sources, as suggested in \cite{hug}.
Figure \ref{redshi_evol} shows a color-color diagram for the HeViCS sources that have an optical counterpart at 250 \micron\ with a reliability above 80\%, a flux density at 250 $\mu$m above 30\,mJy, and
a magnitude in $r$-band below 22.2. We also construct a subsample of $\sim$ 2000 sources with a reliable flux estimation at all SPIRE wavelengths. We divided this subsample in redshift bins of $\Delta z$ = 0.1 to have
a comparable number of sources in each bin, except for the highest ones, where
we are limited by small-number statistics.
The sources at lower redshift span a large region of the diagram, showing an overdensity in
the region with $F_{250}/F_{350}$ $>$ 1.7.
The sources at higher redshift mostly populate the region with $F_{250}/F_{350}$ $<$ 1.7. However, because of the range in temperature and emissivity, this region of the diagram is also heavily
populated by sources at all redshifts. 

The shift of the distribution of sources to lower $F_{250}/F_{350}$ at increasing redshift can be explained by the following considerations:
at low redshifts, the SED of an average galaxy peaks at wavelengths below 250 \micron. The $F_{250}/F_{350}$ ratio appears high and the $F_{500}/F_{350}$ ratio appears low, as
both ratios sample the Rayleigh-Jeans side of the SED. As redshift increases, this peak moves beyond the 250 \micron\ band. Thus, the $F_{250}/F_{350}$ ratio decreases as the bands shift with redshift into
the region near the peak of the SED and onto the Wien side of the SED, and the $F_{500}/F_{350}$ ratio increases, as the 500 \micron\ band approach the peak of the SED.

Most of the sources occupy a region with $F_{250}/F_{350}$ $>$ 0.6 and $F_{500}/F_{350}$ $<$ 1. This is because sources at lower redshift are on average detected at lower SPIRE wavelengths, whereas higher redshift sources have statistically higher flux values at longer wavelengths \cite[see for example][and their
analysis]{hug}. To better investigate this hypothesis in Fig. \ref{redshi_evol}, we showed a reference curve for a modified black body with $\beta$ = 2, 10 $\le T \le$ 30 and a redshift equal to the average of the bin. Fig. \ref{redshi_evol} show that at fixed temperature sources move towards the top left side of the diagram. However, increasing the redshift shows a shift of the curve towards the same direction, reinforcing the idea that increasing the redshift on average sources tend to occupy the region in the top left of the diagram \citep[as shown also in ][]{amb,hug}.

From the color-color diagram, we can infer statistical information about the
redshift distribution of our sources. 
Three different regions can be defined: 
a region with $F_{250}/F_{350}$ $<$ 0.8; a region with 0.8 $<$ $F_{250}/F_{350}$ $<$ 1.7; and a region with $F_{250}/F_{350}$ $>$ 1.7. For the range of temperature and $\beta$ parameters
chosen, the region with lower $F_{250}/F_{350}$ has on average redshift $z \ge$
2 with a higher $F_{500}/F_{350}$ because of the increased emission redshifted
to 500 $\mu$m. In the subset of our sample with SDSS counterparts, the 
number of sources
with these values of the ratios is negligible. The region with 0.8 $<$ $F_{250}/F_{350}$ $<$ 1.7 is the most populated, because there is a large set of
temperature, $\beta$, and redshifts that fall in this range.
The region with $F_{250}/F_{350}$ $>$ 1.7 on average is populated by sources at $z <
1$. The FIR sources exhibit a variation with redshift that follows the trend from bottom right to top left along a curve with increasing
redshift. 

%-----------------------------------------------------------------------------------------------------------
%-----------------------------------------------------------------------------------------------------------

\section{Conclusions} \label{conc}

In this paper, we present three point source catalogs selected at 250, 350, and 500 \micron\
extracted from the HeViCS survey \citep{dav} over an area of $\sim$ 55\,deg$^2$. The FWHM and flux density of the sources have been estimated using the implementation in HIPE 9.0.0 \citep{ott} of \bnd, a timeline-based point source fitter. Then we subtracted the sources from the timeline data using the same Gaussian function that we obtained in the fit.  This subtraction step allowed us to detect objects that are fainter or hidden by other close sources, improving the robustness of our algorithm in terms of identifying sources.

We performed various tests to verify the goodness, reliability, and flux accuracy of the
catalogs. From these analyses, we found that the number of false detections due to background variations
or glitches is negligible in the region selected for the
source extraction. The SPIRE catalogs are about 95\% complete at 40\,mJy, but the completeness decreases at longer wavelengths at lower fluxes. We find a completeness of 88\%, 84\%, and 82\% at 250, 350, and 500 \micron at 30 mJy, while the completeness drops to 72\%, 61\%, and 57\% at 250, 350, and 500 \micron at 20 mJy, confirming that the confusion noise becomes relevant both for flux estimation and for source detection at this threshold. 

We analyze the number counts in the four HeViCS fields, and the results are consistent at
each flux value, in particular for faint fluxes, indicating that Virgo cluster galaxies have a negligible contribution to the total number counts. Another implication of these results is that the Virgo cluster does not contain a significant population infrared-bright, optically-faint of galaxies.

We find an increase of the slope in the number counts at $F \la$ 200\,mJy at all 
SPIRE wavelengths, indicating a strong evolution in luminosities for these galaxies, which agrees with previous studies. Comparison of the number counts at different wavelengths shows that the slope of the faint
end steepens going from 250\,\micron\ to longer wavelengths. Below 40 mJy, the number counts flatten and decrease because of the confusion. This increase in slope might be
due to the presence of actively star-forming galaxies at $z \ge$ 1.5 that are passive at low redshift. 
No model is able to reproduce the bright end of the distribution at all wavelengths, which shows the importance of the Rayleigh-Jeans side of the SED in characterizing the evolution of FIR galaxies
population and the importance of the {\it Herschel} SPIRE instrument that investigates this region. Nevertheless, there is a
relatively small number of sources at the bright end of the number counts, thus precluding definitive conclusions. A point to be better addressed in future modeling efforts is the impact of gravitational lensing on the number
counts, which is mostly relevant at 500 $\mu$m, and the possibility of an evolving SEDs for FIR galaxies with redshift, as found by \cite{magnelli11} and others.

We cross-correlated the SPIRE catalogs with the SDSS DR10 survey
\citep{ahn} and found optical counterparts for a subsample of the HeViCS
sources. For sources with both photometric and spectroscopic redshift
determinations, most sources lie at $z <$ 0.5, with an average
redshift of $z \sim$ 0.3 $\pm$ 0.22 and 0.25 (16-84 percentile).

We divided the catalog into sources with and without optical counterparts and estimated the contribution at each SPIRE wavelength to the total number counts from the low redshift component.
Above 100 mJy, the number of counts is dominated by the component with the optical counterpart (Fig. \ref{nc_red}), which forms the bulk of the bright end of the distribution. Sources with
no optical counterpart underwent a more rapid evolution with respect to the sources with SDSS counterpart.

We then constructed a color-color diagram of the sources with a known redshift. Most of the
sources occupy a region with $F_{250}/F_{350} >$ 0.6 and $F_{500}/F_{350} <$ 1.
We show that most of our sources are  consistent with the chosen ranges of temperature, $\beta$, and $z$. However, this variation must be considered in a statistical sense, because sources with
a lower redshift populate larger regions with an overdensity towards the bottom
right corner of the diagram ($F_{250}/F_{350} >$ 1.7), and sources with high
redshift occupy the top left region of the diagram ($F_{250}/F_{350} <$ 0.8), as shown in Fig. \ref{redshi_evol}. For sources with known redshift, their position in the SPIRE color-color diagram tends to move towards the top-left corner as redshift increases.

\begin{acknowledgements}
We warmly thank the referee for his/her constructive comments and suggestions. We would like to thank D. Munro for freely distributing his Yorick programming language (available at \texttt{http://www.maumae.net/yorick/doc/index.html}).
S. B., L. H., S. Z., S. di S. A. are supported through the ASI-INAF agreement I/016/07/0 and I/009/10/0.
C. P. was also supported by PRIN-INAF 2009/11 grant (extended to 2012).
C.P. acknowledges support from the Science and Technology Foundation (FCT, Portugal) through the Postdoctoral Fellowship SFRH/BPD/90559/2012, PEst-OE/FIS/UI2751/2014, and PTDC/FIS-AST/2194/2012.

Funding for SDSS-III has been provided by the Alfred P. Sloan Foundation, the Participating Institutions, the National Science Foundation, and the U.S. Department of Energy Office of Science. The SDSS-III web site is http://www.sdss3.org/.

SDSS-III is managed by the Astrophysical Research Consortium for the Participating Institutions of the SDSS-III Collaboration including the University of Arizona, the Brazilian Participation Group, Brookhaven National Laboratory, Carnegie Mellon University, University of Florida, the French Participation Group, the German Participation Group, Harvard University, the Instituto de Astrofisica de Canarias, the Michigan State/Notre Dame/JINA Participation Group, Johns Hopkins University, Lawrence Berkeley National Laboratory, Max Planck Institute for Astrophysics, Max Planck Institute for Extraterrestrial Physics, New Mexico State University, New York University, Ohio State University, Pennsylvania State University, University of Portsmouth, Princeton University, the Spanish Participation Group, University of Tokyo, University of Utah, Vanderbilt University, University of Virginia, University of Washington, and Yale University. 
\end{acknowledgements}
   
%\appendix
%\section{Notes on individual objects}
%%%%%%%%%%%%%%%%%%%%%%%%%%%

\end{document}